\documentclass[twocolumn]{aastex631}

\usepackage{wrapfig}
\usepackage{lipsum}

\usepackage{makecell}
\usepackage{orcidlink}
\let\tablenum\relax
\usepackage{siunitx}

\begin{document}

\title{How Meteor Showers Can Guide the Search for Long Period Comets}

\author{Samantha Hemmelgarn\orcidlink{0009-0006-1160-3829}}\footnote{Corresponding Author: shemmelgarn@lowell.edu}
\affiliation{Lowell Observatory, 1400 West Mars Hill Road, Flagstaff, AZ 86001, USA}
\affiliation{Department of Astronomy and Planetary Science, Northern Arizona University \\
P.O. Box 6010, Flagstaff, AZ 86001, USA}

\author{Nicholas Moskovitz}
\affiliation{Lowell Observatory, 1400 West Mars Hill Road, Flagstaff, AZ 86001, USA}

\author{Stuart Pilorz}
\affiliation{SETI Institute, 339 Bernardo Ave, Mountain View, CA 94043, USA}

\author{Peter Jenniskens}
\affiliation{SETI Institute, 339 Bernardo Ave, Mountain View, CA 94043, USA}       

\begin{abstract}
With orbital periods longer than 200 years, most long-period comets (LPCs) remain undiscovered until they are in-bound towards perihelion. The comets that pass close to Earth's orbit are Potentially Hazardous Objects (PHOs). Those with orbital periods up to $\sim$4000 years tend to have passed close to Earth's orbit in a previous orbit and produced a meteoroid stream dense enough to be detected at Earth as a meteor shower. In anticipation of Rubin Observatory's Legacy Survey of Space and Time (LSST), we investigate how these meteor showers can guide dedicated searches for their parent comets. Assuming search parameters informed by LSST, we calculated where the 17 known parent bodies of long-period comet meteor showers would have been discovered based on a cloud of synthetic comets generated from the shower properties as measured at Earth.  We find that the synthetic comets predict the on-sky location of the parent comets at the time of their discovery. The parent comet's location on average would have been 1.51° \textpm1.19° from a line fit through the synthetic comet cloud. The difference between the heliocentric distance of the parent and mean heliocentric distance of synthetic comets on the line was 2.09 \textpm1.89 au for comets with unknown absolute nuclear magnitudes and 0.96 \textpm0.80 au for comets with known absolute nuclear magnitudes. We applied this method to the $\sigma$-Hydrids, the proposed meteor shower of Comet Nishimura, and found that it successfully matched the pre-covery location of this comet 8 months prior to Nishimura's discovery.
\end{abstract}

\section{Introduction} \label{sec:intro}

Meteor showers are caused by streams of meteoroids that move in orbits similar to that of their parent comet. As comets approach perihelion, heat from the Sun warms their surface \citep{comet_nuclei}. Sublimation of gases drags solid particles (the meteoroids) off the nucleus and ejects them into independent orbits, similar to that of the comet because ejection speeds are much smaller than the velocity of the comet around the Sun \citep{comet_nuclei, jenniskens2006, 2p_encke}. The meteoroids will follow slightly different trajectories based on their ejection velocity and the individual grain properties that affect the influence of solar radiation pressure (e.g. size, density, albedo). In returning at different times, these meteoroids disperse along the comet orbit, creating a stream.  Over time, this dispersion will typically increase due to gravitational and non-gravitational perturbations \citep{jenniskens2006, ye2016}. 

Meteor showers sample a subset of all meteoroids in the stream, i.e. only those that intersect Earth's orbit and impact Earth. Earth's intersection with meteoroid streams produce the meteor showers we see in the night sky. A streak of light occurs as meteoric matter is heated and ablated high up in the atmosphere. From that streak, the meteoroid speed and direction of motion can be measured to determine the pre-impact orbit in space. 

The population of long-period comets (LPCs) include a group that have detectable meteoroid streams: a trail of bread crumbs that indicate the comet's presence. LPCs have orbital periods larger than 200 years, too long to resonate with the motion of the planets. Those comets that have orbital periods ($P$) shorter than about 4000 years are known to produce detectable meteor showers at Earth \citep{ms_lpc}. Comets in longer orbits will produce meteoroid streams which become too diluted to be detected as meteors on Earth. 

In recent years, large networks of low-light video camera arrays have been built to map the orbits of long-period comet meteoroid streams. Observations with the Cameras for All-sky Meteor Surveillance (CAMS) network were initiated in 2010 and since then, these camera networks have constrained roughly 3 million meteor orbits \citep{cams, gmn}. The IAU currently recognizes 110 confirmed meteor showers and another 823 awaiting confirmation. 513 showers are described in \citet{Jenniskens2023}, half of which are caused by LPCs. Only 17 of these have known or suspected parent bodies.

A search strategy to find LPCs based on observed meteor shower orbits at Earth was put forth in \cite{JENNISKENS2020104829}. There, the authors used a detected outburst of the 15-Bootids detected by CAMS to constrain a search region on sky for the parent comet by projecting the median keplerian orbital elements of the outburst to potential positions of the parent comet onto the sky. This approach can help find LPCs earlier than in routine searches, which would provide extra warning time before an imminent impact.

Potentially hazardous objects (PHOs) are those comet or asteroid near-Earth objects (NEOs) that pose an impact hazard to Earth. PHOs have absolute magnitudes ($H$) $<$ 22 and a minimum orbit intersection distance (MOID) $<$ 0.05 au. In 1998, NASA set out to catalog all NEOs with a diameter larger than 1 km. Within 13 years, 90\% of the NEO's on short orbits had been discovered and catalogued \citep{neo_results}. Not yet discovered are many of the NEO LPCs, which return to the inner solar system infrequently. The rate of impacts from LPCs could be up to 6\% of all impacts on Earth \citep{QUINTANA2019176}. 

The destructive nature of a single LPC impact is strong motivation to be proactive about the hazards posed by them. Most comets are $\geq$1 km in diameter. The impact of an LPC as small as 1 km can have global consequences because of their high impact velocities. Cometary impact velocities can be as high as $\qty{72}{\km\per\s}$ and are typically $\qty{\sim50}{\km\per\s}$ \citep{Marsden_ihca, Weissman_2006, Nuth_2018}. A 1 km comet of average density, $\qty{0.6}{\g\per\cm\cubed}$  \citep{Weissman_2006, Nuth_2018}, travelling at $\qty{50}{\km\per\s}$ would impact the Earth with roughly the energy of 750,000 megatons of TNT. An impact of this scale could lead to global cooling and ozone layer loss \citep{Toon_ihca}. 

The probability of such an impact on any given perihelion passage remains low, on the order of \num{e-9} \citep{Marsden_ihca} for the whole population. However, among the total population of Oort Cloud comets $\geq$ 1 km with perihelion distances less than 5 AU estimated to be in the range of \num{e12} objects \citep{BOE2019252}, there are $\sim$1000 LPCs that could strike the Earth on their next perihelion passage. $\sim$200 of these have orbital periods small enough to have detectable meteoroid streams.

At present, defending against a potential LPC impactor is made difficult due to the relatively short warning time between discovery and impact \citep{Morrison2006, Nuth_2018}. LPCs are typically discovered around 5 au from the Sun, which only provides up to 2 years warning time to mitigate the impact risk \citep{Gritzner_2006}. In the near future, new technologies will push that detection limit, with the wide-field Legacy Survey of Space and Time (LSST) expecting to see first light in 2025. The potential to discover these comets further out in the solar system will become greater, providing more warning time in the case of a potential impactor. 

If meteor showers could guide the search for LPCs, then dedicated searches could take advantage of these technologies to discover them. The LSST, for example, proposes to take 1000 images of the same patch of sky in a 10-year period \citep{lsst_science_book} and will have the ability to see objects up to a limiting magnitude of 24.5 per exposure \citep{Ivezic_2019}. This limit could be pushed several magnitudes fainter with image stacks that take into account the known rate of motion of the comet. 

Here, we explore the method of \cite{JENNISKENS2020104829} further to investigate how much warning time is gained by using meteoroid streams to guide dedicated deep searches for approaching long-period comets and how feasible such searches are.

\begin{deluxetable*}{cccccc}
\tablecaption{Orbital elements of the 17 meteor showers with known LPC parent bodies. \label{table1}}
\savetablenum{1}
\tablehead{\colhead{Shower/Parent} & \colhead{q} & \colhead{e} & \colhead{i} & \colhead{$\omega$} & \colhead{$\Omega$} \\ 
\colhead{} & \colhead{(au)} & \colhead{} & \colhead{(°)} & \colhead{(°)} & \colhead{(°)} } 
\startdata
$\#6$ & 0.921 \textpm0.016 & 0.954 \textpm0.162 & 79.5 \textpm2.5 & 214.1 \textpm4.3 & 32.3 \textpm1.2 \\
\hspace{2cm}C/1861 G1 & 0.921 & 0.983 & 79.8 & 213.4 & 31.9 \\
$\#23$ & 0.784 \textpm0.058 & 0.953 \textpm0.166 & 170.8 \textpm1.9 & 235.8 \textpm9.9 & 205.2 \textpm7.4 \\
\hspace{2cm}C/1987 B1* & 0.870 & 0.996 & 172.2 & 200.4 & 176.0 \\
$\#145$ & 1.000 \textpm0.005 & 0.953 \textpm0.124 & 74.4 \textpm2.2 & 191.3 \textpm3.5 & 50.0 \textpm1.6 \\
\hspace{2cm}C/1983 H1 & 0.991 & 0.990 & 73.25 & 192.9 & 49.10 \\
$\#175$ & 0.557 \textpm0.056 & 0.964 \textpm0.102 & 149.1 \textpm1.8 & 265.6 \textpm8.3 & 113.3 \textpm10.6 \\
\hspace{2cm}C/1979 Y1 & 0.545 & 0.988 & 148.6 & 257.6 & 103.2 \\
$\#176$ & 0.991 \textpm0.026 & 0.925 \textpm0.198 & 83.6 \textpm6.4 & 341.8 \textpm9.2 & 311.8 \textpm8.4 \\
\hspace{2cm}C/2015 D4* & 0.862 & 0.989 & 77.3 & 314.7 & 305.8 \\
$\#191$ & 0.952 \textpm0.021 & 0.958 \textpm0.19 & 132.4 \textpm3.4 & 28.8 \textpm5.5 & 317.4 \textpm9.3 \\
\hspace{2cm}C/1852 K1 & 0.905 & 1.0 & 131.1 & 37.2 & 319.3 \\
$\#206$ & 0.677 \textpm0.037 & 0.969 \textpm0.094 & 148 \textpm2.1 & 109.4 \textpm6.1 & 158.9 \textpm7.4 \\
\hspace{2cm}C/1911 N1 & 0.684 & 0.996 & 148.4 & 110.4 & 158.7 \\
$\#410$ & 0.920 \textpm0.018 & 0.998 \textpm0.111 & 178.2 \textpm0.9 & 143.8 \textpm4.1 & 90.9 \textpm2.2 \\
\hspace{2cm}C/1864 N1 & 0.909 & 0.996 & 178.1 & 151.6 & 97.7 \\
$\#428$ & 0.626 \textpm0.062 & 0.966 \textpm0.081 & 149.3 \textpm3.0 & 104.8 \textpm8.2 & 275.1 \textpm14.8 \\
\hspace{2cm}C/1846 J1 & 0.634 & 0.990 & 150.7 & 99.7 & 264.0 \\
$\#502$ & 0.796 \textpm0.029 & 0.960 \textpm0.104 & 152.8 \textpm3.0 & 127.4 \textpm5.4 & 256.2 \textpm5.6 \\
\hspace{2cm}C/1961 T1* & 0.681 & 0.992 & 155.7 & 126.6 & 247.4 \\
$\#512$ & 0.988 \textpm0.005 & 0.902 \textpm0.141 & 106.2 \textpm2.9 & 2.2 \textpm7.5 & 45.3 \textpm5.4 \\
\hspace{2cm}C/1879 M1 & 0.896 & 1.0 & 107.0 & 3.74 & 47.5 \\
$\#524$ & 0.916 \textpm0.011 & 0.961 \textpm0.11 & 115.4 \textpm2.2 & 147.4 \textpm3.2 & 214.4 \textpm1.4 \\
\hspace{2cm}C/1975 T2 & 0.838 & 0.985 & 118.2 & 152.0 & 216.8 \\
$\#531$ & 0.985 \textpm0.008 & 0.941 \textpm0.123 & 123.2 \textpm2.0 & 198.1 \textpm3.4 & 48.7 \textpm2.9 \\
\hspace{2cm}C/1853 G1* & 0.909 & 0.989 & 122.2 & 199.2 & 43.0 \\
$\#533$ & 0.851 \textpm0.063 & 0.958 \textpm0.145 & 170.9 \textpm2.8 & 312.1 \textpm10.4 & 291.9 \textpm10.8 \\
\hspace{2cm}C/1964 N1 & 0.822 & 0.985 & 171.9 & 290.8 & 269.9 \\
$\#535$ & 0.514 \textpm0.031 & 0.999 \textpm0.07 & 138.5 \textpm1.7 & 89.1 \textpm5.6 & 312.9 \textpm2.3 \\
\hspace{2cm}C/1939 H1 & 0.528 & 0.991 & 138.1 & 89.2 & 312.3 \\
$\#545$ & 0.724 \textpm0.015 & 0.966 \textpm0.053 & 97.5 \textpm1.0 & 245 \textpm2.4 & 149.4 \textpm1.0 \\
\hspace{2cm}C/1871 V1* & 0.691 & 0.995 & 98.3 & 242.9 & 148.9 \\
$\#705$ & 0.781 \textpm0.026 & 0.950 \textpm0.084 & 103.8 \textpm1.8 & 121.8 \textpm4.4 & 166.3 \textpm3.0 \\
\hspace{2cm}C/2002 Y1 & 0.714 & 0.997 & 103.8 & 128.8 & 166.3 \\
\enddata
\tablecomments{Median observed shower orbital elements with dispersions and associated parent body as found in \cite{ms_lpc}. Asterisk (*) indicates uncertain parent/shower associations. Comet elements from JPL Horizons Small-Body Database.}
\end{deluxetable*}

\section{Methods} \label{sec:methods}
To investigate the use of meteor shower data in guiding searches for their LPC parents, we investigated how the 17 known long-period comet parent bodies (Table \ref{table1}) would have been detected in a dedicated search.

Note that not all comet shower associations are certain. Parent body associations to meteor showers are typically determined by comparing median orbits by a method known as the D-criterion of \cite{s&h_d_criterion}. This criterion calculates how dissimilar two orbits are. \cite{DRUMMOND1981545, JOPEK1993603, Valsecchi_MSI_1_1999, JENNISKENS200813, Jopek2008, Rozek_2011, Rudawska_2015} have all expanded on \cite{s&h_d_criterion} to provide additional methods for determining parent/shower association. However, dynamical modelling of debris ejected from the shower's potential parent is a better method to confirm these associations \citep[e.g.][]{Segon_2017}. Here, we refer to those past studies, summarized in \citep{Jenniskens2023}, and evaluate the likelihood of the proposed associations from our analysis.

\subsection{Range of possible comet positions on sky} \label{sec:2.1}

For each case, we created a cloud of synthetic comets based solely on the observed dispersion of the meteor shower orbital elements measured at Earth (Table \ref{table1}). A ``search region", for the purposes of this analysis, is defined as the solid angle subtended by the cloud of synthetic comets. We defined a heliocentric distance ($r$) range that would put the synthetic comets just beyond current survey brightness limits ($V\sim22$), yet consistent with expected LSST capabilities ($V<25$), and calculated where the known comet would have been detected relative to the cloud had LSST been operating. We fit a line through the cloud of synthetic comets to conduct our analysis. We then determined how far from the center of the cloud the comet would have been located, how far it would have been from Earth's orbit, and how much extra warning time would have been provided. 

\begin{figure}[h]
    \centering
    \includegraphics[width=0.5\textwidth]{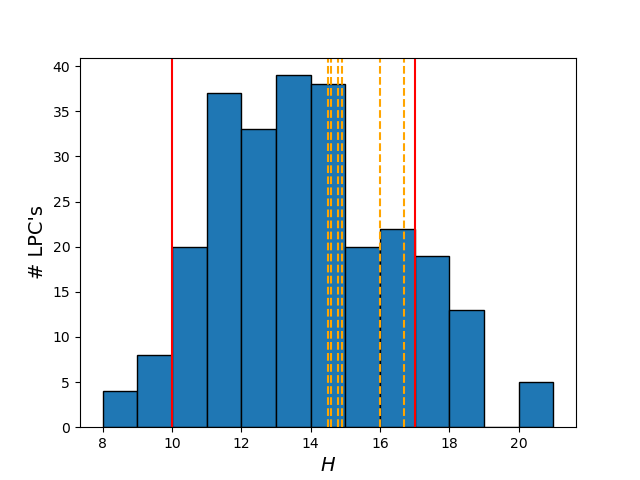}
    \caption{Distribution of $H$ values for long period comets from JPL's Small Body Database. Solid red lines denote assumed cutoff values for comets with undefined bare nucleus magnitudes. Dashed orange lines indicate nominal $H$ values for the 6 comets in our sample which have this property constrained.}
    \label{m2_dist}
\end{figure}

We used the nominal $H$ and its corresponding heliocentric distance range for 6 comets with known $H$ values. For those that did not have known $H$ values, we kept this as a free parameter and defined the suitable range of $r$ for each integer absolute magnitude between $10 \le H\le 17$ (Figure \ref{m2_dist}). 

We then determined in which part of the orbit the comet approached perihelion and was within heliocentric distances corresponding to visual magnitudes ($V$) between $22 \le V \le 25$. For example, a comet with an $H=14$ would have $6.224 \le r \le 11.887$ where $V$ was within this specific range. Based on the time of perihelion, we determined the date on which our $H=14$ parent comet was at $r=6.224$, $r=7$, $r=8$, etc., up to $r=11.887$. For each date (i.e. heliocentric distance), we created an ephemeris and projected onto the sky only the synthetic comets that were on LPC-like orbits ($200 \le P \le 4000$) and within our observability constraints, along with the position of the parent comet on that date. In all, we completed 850 simulations by following this process across 17 meteor showers. 

\subsection{Software Tools} \label{sec:2.2}
Our calculation of projected comet positions on sky used the Python programming language and the open-source orbit computation program \fontfamily{cmtt}\selectfont
OpenOrb \fontfamily{cmr}\selectfont \citep{OpenOrb}. We used \fontfamily{cmtt}\selectfont OpenOrb \fontfamily{cmr}\selectfont to calculate projected sky positions (RA, Dec) on a given date (and thus position of Earth in its orbit) for an array of synthetic comets, derived from a uniform sampling of each meteoroid stream's measured dispersions. \fontfamily{cmtt}\selectfont OpenOrb \fontfamily{cmr}\selectfont has functions that include orbit determination, mass determination, ephemeris generation and orbit propagation. Ephemeris generation was the sole function that we used. The function reads an orbit file (that must be of file extension .orb or .des, .des used here) containing orbital elements, absolute magnitude, date, and coordinate type (CAR: cartesian, KEP: keplerian, COT: cometary true anomaly, etc.). The IAU observatory code is a parameter that is specified during ephemeris generation to define the topocentric coordinates for the observing location. Obscode G37, corresponding to the Lowell Discovery Telescope in Happy Jack, AZ, was used in our analysis.   

\subsection{Comet observability: defining the sweet-spot} \label{sec:2.3}
Our ideal search area is when the comet is at opposition, approaching perihelion (true anomaly $>180$°), and where it has an apparent brightness fainter than achieved in routine observations by ongoing asteroid surveys ($V\sim22$), but still within reasonable limits for current targeted or future wide-field surveys ($V<25$). 

Comets brighter than $V=22$ would have already been discovered by existing surveys such as Pan-STARRS or the Zwicky Transient Facility \citep{Flaugher_2015, Bellm_2019, chambers2019panstarrs1}. The upper limit on $V$ corresponds to the approximate single image depth of LSST \citep{Ivezic_2019}. Opposition is where the comet appears brightest as seen from Earth. Comets are typically discovered while they are approaching perihelion because they become brighter as they get closer to the Sun. We exclude outbound (post-perihelion) comets because they would already have been discovered when their apparent brightness was high due to close proximity.  

Comet brightness is due to the size of the nucleus, but also the rate at which dust and gas is ejected. Comets are known to begin sublimating water ice at $\sim$5 au \citep{Whipple_1950, Jewitt_2007}, but can also exhibit activity at distances as far out as $\sim$35 au where activity is driven by sublimation of volatiles other than water or by other physical processes of ejection \citep{Jewitt_2021}.

Because the behavior of the comet can vary, we adopt a conservative approach by assuming the comet is inactive at discovery. The assumption of a bare comet nucleus gives us conservative estimates of the comet's $V$ at any given $r$, as any increase in the comet's activity would make it brighter and easier to find. The $V$ of an inactive comet is derived from $H$. The $H$ of the comet nucleus is a parameter passed into \fontfamily{cmtt}\selectfont OpenOrb \fontfamily{cmr}\selectfont that is one of the predominant contributors to the variance of outcomes in our simulations. 

Only six of the 17 comets we modeled have observed $H$'s with specified uncertainties. Five of these six comets had $H$'s listed in the JPL Small Body Database. Here, we use the term $H$ interchangeably with the nuclear magnitude $M2$, as is often used for comets. The $H$ values for comets with measured nuclear magnitudes are listed in Table \ref{table2}. The absolute magnitude for C/1983 H1 (Iras-Araki-Alcock) was published by \cite{GROUSSIN2010904}. 

To determine the suitable range of $H$ for comets with no known values, the JPL Small Body Database was searched for hyperbolic or parabolic comets with known $H$ values. From these 259 comets, we obtained a distribution of nuclear magnitudes (Figure \ref{m2_dist}) to inform a plausible range of H. Just over 82$\%$ of $H$'s in this distribution fall into the 10 to 17 magnitude range, bounds that are then used to define the range of $H$ for showers with unconstrained nuclear magnitudes. Cometary albedos fall within a typical range of 0.04 to 0.06 \citep{knight2023physical}, thus $H<17$ corresponds to diameters $>\sim1$ km, and would be considered potentially hazardous. The drop off in numbers above H = 17 is thought to be real, with relatively few comets known to have a diameter much less than 1 km.  

Assuming a viewing geometry that puts the comet at opposition, we solved for the range of $r$ which would put the comet within our brightness limits \citep{comet_nuclei_properties}:

\begin{equation}
\label{eqn1}
V = H + 5\log_{10}(r\Delta) - 2.5\log_{10}q(\alpha)
\end{equation}
Where $V$ is visual magnitude, $H$ is the comet's absolute magnitude, $r$ is the comet's heliocentric distance, $\Delta$ is the distance between the object and the Earth, and $q(\alpha$) is the phase integral at solar phase angle $\alpha$. Assuming the comet nucleus is an inactive, Lambertian surface, the phase integral is given by \citep{planet_brightness}:

\begin{equation}
\label{eqn2}
q(\alpha) = \frac{2}{3}\left[\frac{(\pi - \alpha)\cos{\alpha} + \sin{\alpha}}{\pi}\right]
\end{equation}
Because searches would be optimized when the object is at opposition ($\alpha$ = 0°), the phase integral reduces to $\frac{2}{3}$.  $\Delta$ in terms of heliocentric distance becomes $r$ - 1, providing a way to calculate visual magnitudes solely as a function of $H$ and $r$.

\begin{figure}
    \includegraphics[width=0.5\textwidth]{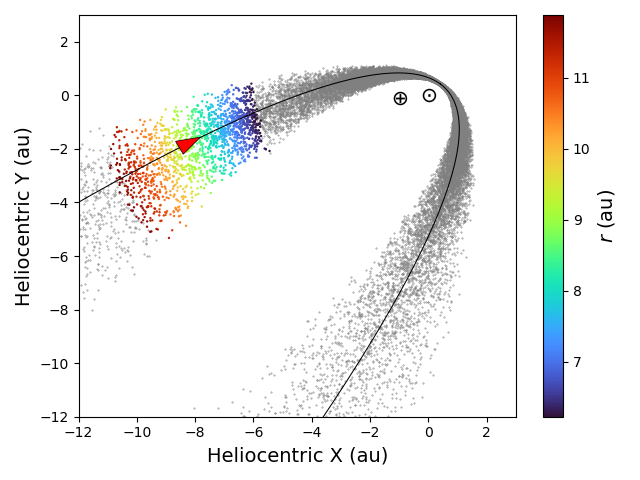}
    \caption{The observability sweet spot for the Aurigids (IAU shower 206) of comet C/1911 N1 (Kiess) for the date of March 27, 1909, when its parent comet was 9 au from the Sun. The red arrow marks the comet position and direction of motion at that time. $\odot$ is the Sun and $\oplus$ the Earth. Gray circles are the synthetic comets on Aurigid-type orbits. The region with colored symbols highlights the heliocentric distance range that would be observable for a comet with $H=14$.}
    \centering
    \label{206_orbit}
\end{figure}

\subsection{On sky ephemeris} 
\label{sec:2.4}

We created synthetic comets by generating 1000 random clones for each 1° of true anomaly, resulting in 360,000 random clones for each shower. Each clone is pulled from a uniform random sample of the shower’s five Keplerian orbital elements. A uniform distribution inflates our calculated search regions but provides conservative estimates on where to begin searching, a method which can be refined with future work. We filtered out synthetic comets that were not on LPC-like orbits and fell outside of the observability constraints. 

We computed an ephemeris for each synthetic comet using \fontfamily{cmtt}\selectfont
OpenOrb \fontfamily{cmr}\selectfont (utilizing the COT coordinate type). For the purpose of this test, observatory code G37 (Lowell Discovery Telescope) was used. \fontfamily{cmtt}\selectfont
OpenOrb \fontfamily{cmr}\selectfont returns epoch J2000 RA and Dec positions in the topocentric equatorial coordinate system.  

\begin{figure}
    \centering
    \includegraphics[width=0.5\textwidth]{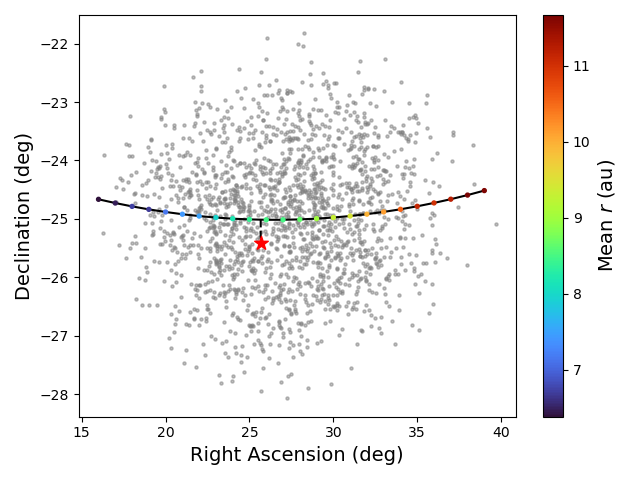}
    \caption{Case study for the shower associated with comet C/1939 H1 (Jurlof-Achmarof-Hassel), the position of which is shown by a red star. Colored dots on the line represent the mean $r$ of synthetic comets in 1° bins along the line. Its $\delta_l$ is 0.40° and $\delta_r$ is 0.39 au in this simulation.}
    \label{535_8_h_14}
\end{figure}

The dates for ephemeris generation were dictated by the parent comet. The date selected was based on the parent comet's heliocentric distance from the sun on its previous passage through the solar system. We pulled dates from the JPL Small Body Database corresponding to when the parent comet was at heliocentric distances which put its $V$ within the range defined earlier, $22 \le V \le 25$. We generated an ephemeris for each date corresponding to those $r$'s. This approach placed the synthetic comets in what was the parent comet's path on its last perihelion passage. This allows us to constrain the parent comet's location with respect to the synthetic comets. 

Observable synthetic comets with LPC-like orbits, along with the shower's parent comet, were plotted by their position on sky in right ascension and declination. Figure \ref{535_8_h_14} provides an example of a search area on the sky for C/1939 H1 (Jurlof-Achmarof-Hassel). The figure is a visualization of observable synthetic comets (gray circles) simulated using median orbital elements and their dispersions from shower $\#535$, $H=14$, on the date the parent comet was at 8 au (May 16, 1937).

The RA and Dec positions depend on the position of Earth in its orbit and the part of the comet orbit that is selected from the assumed range of brightness. 

To help determine the offset in position between the parent comet and the most likely position of the synthetic comets, we fit a second order polynomial to the RA/Dec positions of the synthetic comets (Figure \ref{535_8_h_14}). This line was oriented along the longest axis of the synthetic comet field, i.e. the RA or Dec coordinate that displayed the greatest range. The line provides a mean representation of synthetic comet positions within the cloud as well as a reference point to compare the position of the parent comet. 

An important fact to underscore is that our modelling is only attempting to show that the shower can serve as a guide to where the comet may be in the sky. It does not take into consideration viewing geometry (e.g. solar elongation), which would determine whether any particular patch of sky can be observed from Earth on a given night.

\begin{deluxetable*}{ccccccccccc}
\tablecaption{Mean simulation results by meteor shower. \label{table2}}
\savetablenum{2}
\tablehead{\colhead{} & \colhead{Shower} & \colhead{H} &\colhead{$\delta_l$ \textpm$\sigma_l$} & \colhead{$\Psi$} & \colhead{$\psi_l$} & \colhead{$\delta_r$ \textpm$\sigma_r$} & \colhead{$\bar{K_s}$} & \colhead{$\bar{K_p}$} & \colhead{$\bar{\theta_s}$} & \colhead{$\bar{\theta_p}$}\\
\colhead{} & \colhead{} & \colhead{} & \colhead{(°)} &  \colhead{(°)} & \colhead{\%} & \colhead{(au)} & \colhead{(\arcsec/min)} & \colhead{(\arcsec/min)} & \colhead{(°)} &  \colhead{(°)}} 
\startdata
Unknown H & 6 & 10-17 & 1.14 \textpm1.39 & $<$4.67 & 24 & 1.43 \textpm1.42 & 0.225 & 0.229 & 150.3 & 153.2\\
Showers & 23* & 10-17 & 2.85 \textpm0.64 & 2.16 & 132 & 2.41 \textpm1.75 & 0.146 & 0.153 & 189.5 & 188.0\\
        & 175 & 10-17 & 1.65 \textpm0.65 & 3.57 & 47 & 2.44 \textpm1.82 & 0.180 & 0.173 & 163.2 & 162.8\\
        & 191 & 10-17 & 3.33 \textpm1.28 & 5.92 & 56 & 2.47 \textpm2.35 & 0.248 & 0.206 & 183.7 & 183.8\\
        & 206 & 10-17 & 0.83 \textpm0.41 & 4.09 & 20 & 2.35 \textpm1.98 & 0.187 & 0.172 & 191.0 & 191.8\\
        & 410 & 10-17 & 0.25 \textpm0.04 & 0.30 & 83 & 1.70 \textpm1.75 & 0.162 & 0.155 & 176.7 & 175.4\\
        & 428 & 10-17 & 1.70 \textpm0.83 & 5.65 & 30 & 2.60 \textpm2.00 & 0.183 & 0.184 & 170.0 & 171.7\\
        & 512 & 10-17 & 1.48 \textpm0.88 & 7.10 & 21 & 2.05 \textpm1.81 & 0.190 & 0.184 & 178.6 & 174.5\\
        & 531* & 10-17 & 1.81 \textpm0.84 & 3.25 & 56 & 2.15 \textpm2.23 & 0.216 & 0.202 & 172.7 & 181.1\\
        & 535 & 10-17 & 0.53 \textpm0.34 & 2.48 & 21 & 2.13 \textpm1.81 & 0.192 & 0.187 & 203.1 & 204.0\\
        & 545* & 10-17 & 1.00 \textpm0.33 & 1.62 & 62 & 1.26 \textpm0.89 & 0.212 & 0.225 & 177.7 & 179.7\\
        \cline{2-11}
        & Mean &       & 1.51 \textpm1.19 & 3.71 & 50 & 2.09 \textpm1.89 & 0.195 & 0.188 & 177.9 & 179.7\\
\hline
Known H & 145 & 14.9 & 0.57 \textpm0.43 & 2.62 & 22 & 0.44 \textpm0.20 & 0.313 & 0.302 & 148.4 & 146.9\\
Showers & 176* & 14.8 & 13.89 \textpm5.06 & 10.60 & 131 & 2.28 \textpm1.57 & -- & 0.253 & -- & 169.1\\
        & 502* & 16.7 & 2.75 \textpm1.23 & 3.68 & 75 & 0.88 \textpm0.51 & 0.312 & 0.301 & 155.0 & 156.0\\
        & 524 & 14.6 & 1.51 \textpm0.51 & 2.09 & 72 & 0.78 \textpm0.55 & 0.203 & 0.188\ & 184.0 & 184.5\\
        & 533 & 16.0 & 2.92 \textpm0.48 & 2.28 & 128 & 0.98 \textpm0.60 & 0.182 & 0.189 & 176.1 & 169.7\\
        & 705 & 14.5 & 1.22 \textpm0.72 & 4.03 & 30 & 1.67 \textpm1.09 & 0.239 & 0.204 & 250.0 & 248.9\\
        \cline{2-11}
        & Mean &      & 1.70 \textpm1.12 & 4.22 & 76 & 0.96 \textpm0.80 & 0.250 & 0.237 & 182.7 & 181.2\\
\hline
Overall Average   &      &      & 1.51 \textpm1.19 & 3.89 & 59 & 2.05 \textpm1.87 & 0.212 & 0.204 & 179.4 & 179.5\\
\enddata
\tablecomments{Average distance of parent comet from the synthetic comet trend line ($\delta_l$), synthetic comet cloud width ($\Psi$), average difference in heliocentric distance between the comet and particles on the line ($\delta_r$), standard deviation ($\sigma_l$ and $\sigma_r$) of those calculations for all 17 showers, the mean non-sidereal rate for synthetic comets ($\bar{K_s}$) and their parents ($\bar{K_p}$), and the direction of motion measured in degrees east of north for synthetic comets ($\bar{\theta_s}$) and their parents ($\bar{\theta_p}$) across all simulations (n=850).  $\Psi$ is the distance from the trend line to the outside of the cloud which encompasses 99.7\% of all synthetic comets. $\psi_l$ presents $\delta_l$ as a percentage of $\Psi$. $\delta_r$ is the difference between the heliocentric distance of the parent comet and the heliocentric distance of synthetic comets in the 1° bin at the point on the line where $\delta_l$ is measured. $\sigma_l$ and $\sigma_r$ measure how widely values of $\delta_l$ and $\delta_r$ are dispersed from the mean. $\bar{K_s}$ and $\bar{\theta_s}$ is the mean rate and angle (measured east of north) for all synthetic comets inside a 3° FOV around the parent comet across all simulations. $\bar{K_p}$ and $\bar{\theta_p}$ is the mean rate and angle of the parent comet pulled from the JPL Small Body Database for the date of each simulation. Asterisk (*) indicates uncertain parent/shower associations.}
\end{deluxetable*}

\begin{figure*}
    \centering
    \includegraphics[width=1\textwidth]{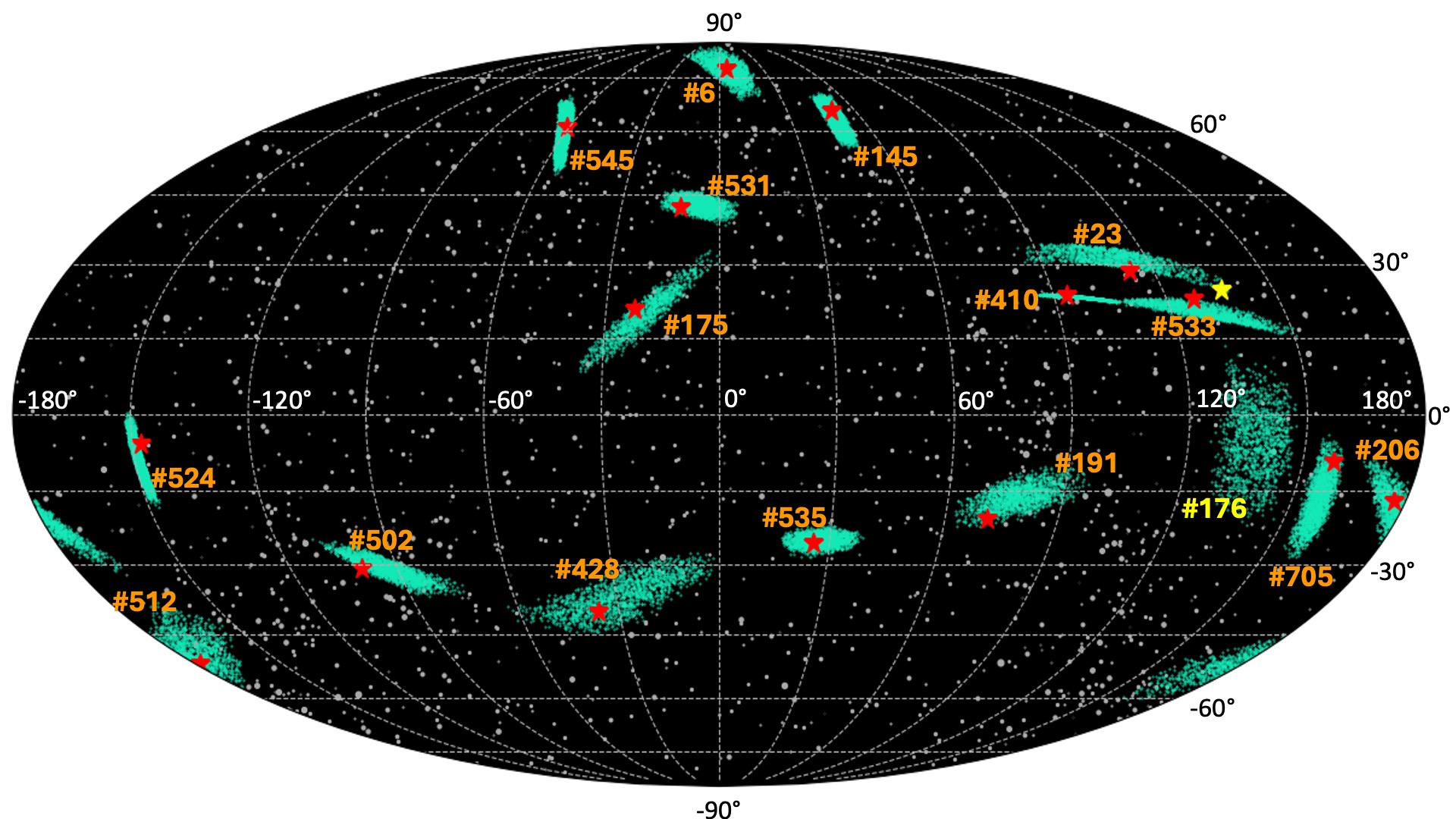}
    \caption{A celestial sphere projecting all 17 synthetic comet clouds, with the position of the parent comet plotted as a red star. Orange text denotes the shower number used to generate the cloud.}
    \label{celestial_sphere}
\end{figure*}

\section{Results} \label{sec:results}

\subsection{Size and shape of the search area} 
\label{sec:3.1}

The dispersions in Table \ref{table1} are the standard deviation of orbital elements for measured meteors in each shower, after removing the effect of nodal precession \citep{Jenniskens2023}. The projected area on sky for a set of synthetic comets is dependent upon these measured dispersions. Figure~\ref{celestial_sphere} is a representation of our 17 showers. Some search areas are diffuse, others are compact. 

The comet position is plotted as red stars. In general, the comet positions were found to be projected inside the cloud of synthetic comets, despite the meteoroid stream only having been sampled in Earth's path. The parent comet of two showers  ($\#23$ and $\#533$, Figure \ref{celestial_sphere}) are found right on the edge of the projected cloud. 

The one clear exception is the case of shower \#176, labeled in yellow, where the proposed parent comet C/2015 D4 projects at a higher declination than the cloud. This is also our most dispersed shower and with a high amount of precession of the elements in Earth's path due to a low inclination of the orbit. The disagreement could mean that C/2015 D4 is not the parent body of shower \#176, or that the stream extends well beyond the part of the stream observed at Earth as a meteor shower.

Differences in brightness between the parent comet and synthetic comets along the comet orbit are attributed to differences in heliocentric distance (Equation \ref{eqn1}). This defines two characteristic uncertainties: one along-orbit responsible for the highest on-sky dispersion (reflecting spread along the comet orbit from variation in $H$), and the other perpendicular to that line (from the perpendicular dispersion of the shower). 

Determining the perpendicular distance from the line to the parent comet ($\delta_l$) gave us a representative uncertainty in on-sky position. The width of the synthetic comet cloud ($\Psi$) is defined as the distance from the line to the outside of the cloud that encompasses 99.7\% (3$\sigma$) of all synthetic comets. $\Psi$ was computed by calculating $\delta_l$ for each synthetic comet and using a cumulative frequency distribution of all distances to determine the 3$\sigma$ width. 3$\sigma$ width is used to exclude outlying synthetic comets whose distance from the line is not representative of the rest of the distribution. We use this to express $\delta_l$ as a percentage of the width of the cloud ($\psi_l$) and provide a more illustrative description of where the parent comet was located with respect to the synthetic comets.

The along-orbit dispersion is a measure of uncertainty caused by the unknown comet brightness. For each 1° bin on the along-orbit trend line, we calculated the mean $r$ of the synthetic comets in that bin. The difference between mean $r$ of the synthetic comets on the line at that point and $r$ of the parent comet ($\delta_r$) suggests a characteristic uncertainty in $r$ that constrained the apparent non-sidereal motion of the object, since non-sidereal motion will correlate with heliocentric distance.  

Table \ref{table2} summarizes calculations of $\delta_l$, $\delta_r$, $\Psi$, and $\psi_l$, completed across all simulations. 
Results for shower \#176 are not included in the calculated averages.

These numbers are also affected by the coordinate system used. One of our showers ($\#6$) had a cloud of synthetic comets that stretched over the northern pole in several of the simulations that were run. A skewed fit line resulted in this situation. After excluding 6 simulations where the synthetic comet cloud stretched over the north pole out of 74 total simulations, the mean $\delta_l$ improved by $0.11$°, $\sigma_D$ improved by $0.19$°, $\delta_r$ increased by $<5\%$, and $\sigma_r$ increased by 0.03 au, while $\Psi$ improved by 0.25° (5.4\%).

\begin{figure*}
    \gridline{\fig{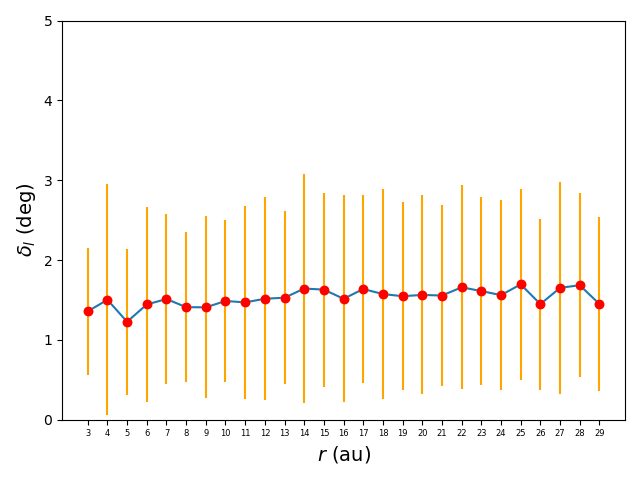}{0.5\textwidth}{(a)}
    \fig{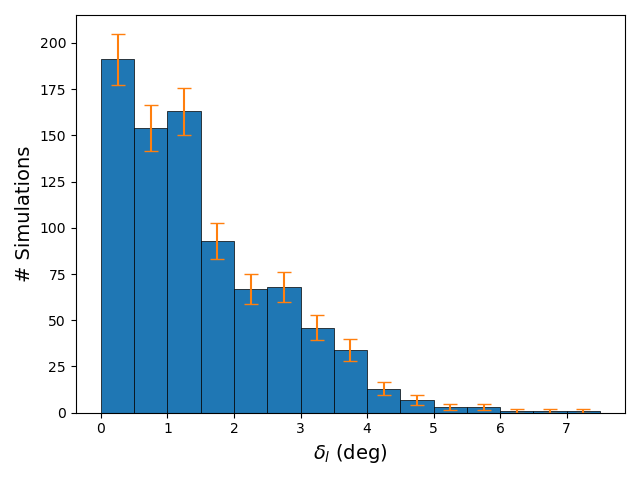}{0.5\textwidth}{(b)}}
    \gridline{\fig{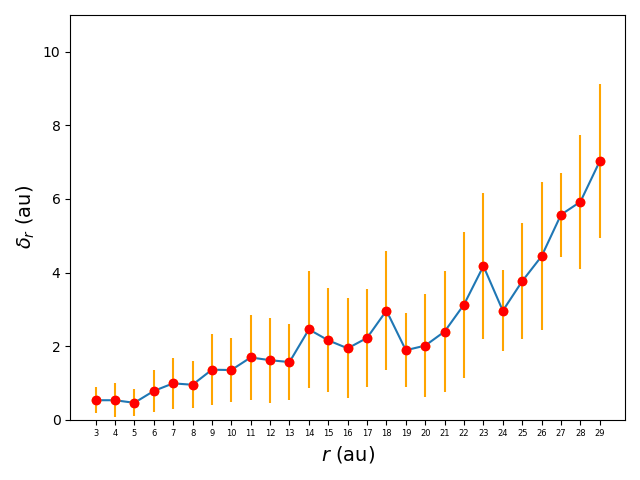}{0.5\textwidth}{(c)}
    \fig{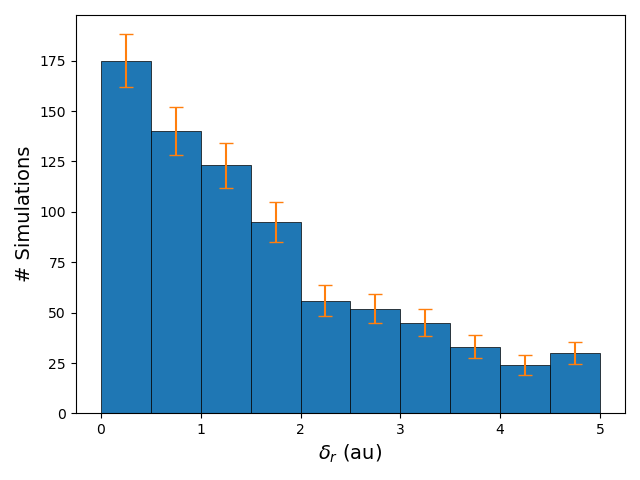}{0.5\textwidth}{(d)}}
    \caption{Line plots (a \& c) show average values of $\delta_l$ and $\delta_r$ with standard deviations for all showers as a function of heliocentric distance.  Histograms (b \& d) provide distributions of $\delta_l$ and $\delta_r$ for all simulations across all showers. Error bars in the line plots are the standard deviations across all simulations. Error bars in the histograms are calculated as $\sqrt{N}$.}
    \label{plots_and_hists}
\end{figure*}

Figure \ref{plots_and_hists} shows $\delta_l$ and $\delta_r$ (a \& c) as a function of heliocentric distance across all 850 simulations. Histograms (b \& d) show distributions of $\delta_l$ and $\delta_r$ across all simulations. 

\subsection{The direction of motion for shift and stack searches} \label{sec:3.2}

\begin{figure}
    \centering
    \gridline{\fig{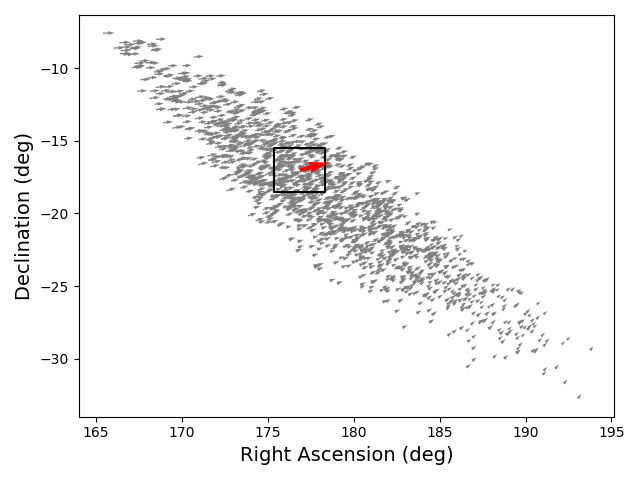}{0.41\textwidth}{}\label{rate_main}}
    \gridline{\fig{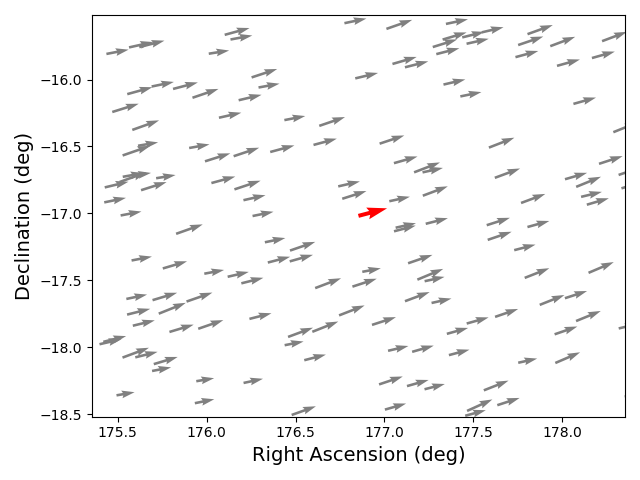}{0.405\textwidth}{}\label{rate_inset}}
    \caption{Top plot shows velocity vectors for each observable synthetic comet in a simulation with shower $\#206$ (the Aurigids) on July 31, 1909. The parent comet's velocity vector is represented with a red arrow. The black box is a 3° FOV region of the search area centered around the parent comet's position and represents the FOV of a wide angle sky survey. Bottom plot is a close up of this region, showing that velocity vectors of synthetic comets are representative of the velocity vector of the parent comet. The mean rate of synthetic comets in this 3° region are 0.144 $\pm$0.016 \arcsec/min and the rate of the parent comet is 0.145 \arcsec/min.}
    \label{shift_stack}
\end{figure}

A shift and stack search strategy \citep{Bernstein_2004, Zhai_2020} involves stacking a sequence of images at a specified rate and angle of motion in an effort to access fainter magnitudes. Shifting and stacking of images can be computationally intensive if bounds on the rate and angle on sky are unknown, as one would need to test all possible parameters (rates and angles).

By comparing the velocity vectors of the synthetic comets to those of the parent comets in our sample, we were able to provide useful bounds to this type of search strategy. For example, LSST will have a 3.5° field of view (FOV) \citep{lsst_science_book}. Mean velocity vectors of the synthetic comets inside a 3° FOV around the parent comet were compared to the parent's velocity vector. Table \ref{table2} summarizes the mean rates ($\bar{K_s}$) and angles ($\bar{\theta_s}$) of the synthetic comets inside the 3° FOV along with the mean rates ($\bar{K_p}$) and angles ($\bar{\theta_p}$) of the parent comets across all simulations. Figure \ref{shift_stack} depicts how the velocity vectors of the synthetic comets inside such a region compare to that of the parent. 

\begin{figure}
    \centering
    \includegraphics[width=0.45\textwidth]{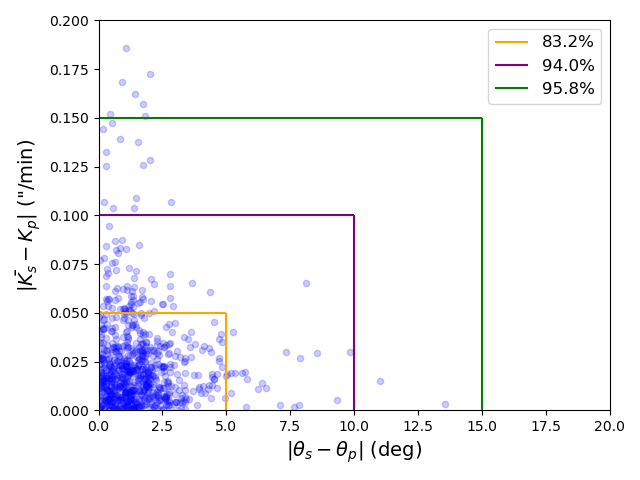}
    \caption{Differences between mean synthetic comet rates and angles and those of the parent comets inside a 3° FOV search region encompassing the parent comet. Legend shows percentage of total simulations (n=850) that fall within bounds of 0.05 \arcsec/min and 5° (yellow), 0.1 \arcsec/min and 10° (purple), and 0.15 \arcsec/min and 15° (green)}
    \label{rate_compare}
\end{figure}

Figure \ref{rate_compare} shows the aggregate of all of the differences between mean rates and angles of synthetic comets and those of the parent comet. $\sim$96\% of the parent comets were traveling at rates \textpm0.15 \arcsec/min and \textpm15° from the mean direction of motion of the synthetic comets within a 3° FOV, defining bounds which can then be used in a shift and stack search for faint objects.

\begin{figure}
    \centering
    \includegraphics[width=0.45\textwidth]{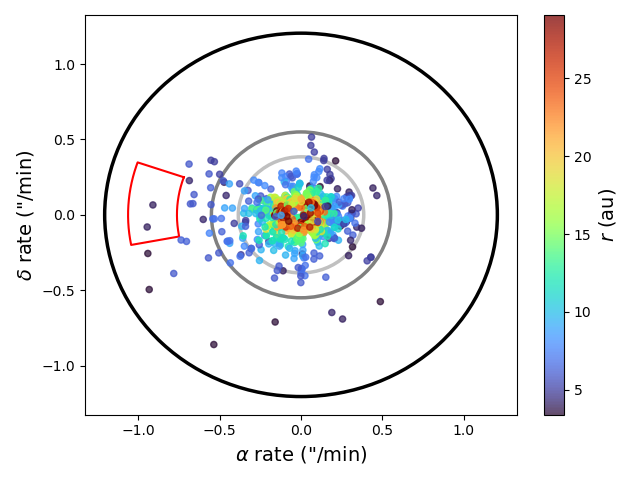}
    \caption{RA/Dec rates (\arcsec/min) for each simulation (n=850) color-coded by $r$. The outer circle represents case 1, the middle circle case 2, and the inner circle case 3. The example wedge sets bounds on the rates and angles for a shift and stack search in the region defined by shower $\#531$ with $H=14$ and $r=4$ au, based on bounds discussed in section \ref{sec:3.2} (±0.15 \arcsec/min and ±15°). These bounds reduce the parameter space by $96.5\%$ for this simulation.}
    \label{rates_scatter}
\end{figure}

We estimate a quantitative reduction in this parameter space by taking the ratio of the number of bound parameters to the total number of (unbound) parameters in dRA/dDec (\arcsec/min) space. We defined the total parameter space by estimating a maximum possible rate for the synthetic comets. This allowed us to encircle all rates and position angles to be tested in a blind shift and stack search (Figure \ref{rates_scatter}). We tested three maximum rate cases to understand their influence on the ratio of bound parameters to total parameters: the maximum rate across all simulations (1.06 \arcsec/min); a maximum rate of 0.400 \arcsec/min, which excludes fast-moving outliers and encompasses 763 ($\sim90\%$) of the mean rates found in the 850 simulations; and the maximum simulated rate at heliocentric distances greater than 15 au (0.235 \arcsec/min). This last test focuses on faint, slow-moving objects where searches would likely be conducted. 341 ($\sim40\%$) of our simulations fell into this category.

For each of these cases, we added another 0.15 \arcsec/min to set a limit on the maximum allowed rate. This defined the total possible parameter space to be searched. For each simulation we then calculated the area in a wedge \textpm15 \arcsec/min and \textpm15° centered on the mean rate of each shower (Figure \ref{rates_scatter}). This wedge represents the reduction in parameter space to be tested when conducting a shift and stack search for a specific parent body.

When the upper bound was based on the maximum rate found in our simulations (first case), we found that the parameter space was reduced by at least 96.4\%. This maximum rate is not necessarily representative of our entire sample because outliers inflate the size of the allowable parameter space, thus deflating the computed ratio. When the upper bound was based on a maximum rate of 0.400 \arcsec/min (second case), we found that the parameter space was reduced by at least 93.4\%. Finally, the parameter space was reduced by at least 92.1\% for simulations at heliocentric distances greater than 15 au (third case). This reduction in parameter space of over 92\% would decrease the computation time by about an order of magnitude when conducting shift and stack searches.

\subsection{The additional warning time and distance from Earth} \label{sec:3.3}
To address the potential for additional impact warning time provided by this method, we backwards integrated our parent comets from their dates of first observation to the date where $V=25$. $V$ was calculated from equation \ref{eqn1} by using the $r$ and $\Delta$ of the object on dates prior to discovery. This allowed us to calculate $V$ of inactive comet nuclei, consistent with an assumption made throughout our modelling. 

The difference between the discovery date and the date where $V=25$ informs the additional warning time provided if a long period comet on a potential Earth intersecting orbit were discovered using this method. Table \ref{table3} provides calculations of the additional years before discovery and the distance from the Earth if these LPCs were discovered at $V=25$ for objects with $10 \le H \le 17$. In that metric, the largest comets ($H=10$) would be discovered further out in the solar system and provide over 12 years additional warning time, while the smallest comets ($H=17$) would still provide over a year to determine and execute mitigation strategies.

\begin{deluxetable}{ccccc}
\tablecaption{Additional warning time provided and distance from Earth at discovery. \label{table3}}
\savetablenum{3}
\tablehead{\colhead{H} & \colhead{} & \colhead{Years Warning} & \colhead{} & \colhead{$\Delta$ (au)}}
\startdata
10 & & 12.6 & & 28.9 \\
11 & & 8.8 & & 23.1 \\
12 & & 6.4 & & 18.1 \\
13 & & 4.6 & & 14.5 \\
14 & & 3.3 & & 11.5 \\
15 & & 2.4 & & 9.2 \\
16 & & 1.7 & & 7.5 \\
17 & & 1.2 & & 5.9 \\
\hline
Mean & & 5.1 & & 14.8
\enddata
\end{deluxetable}

\begin{figure*}
   \centering
    \includegraphics[width=\textwidth]{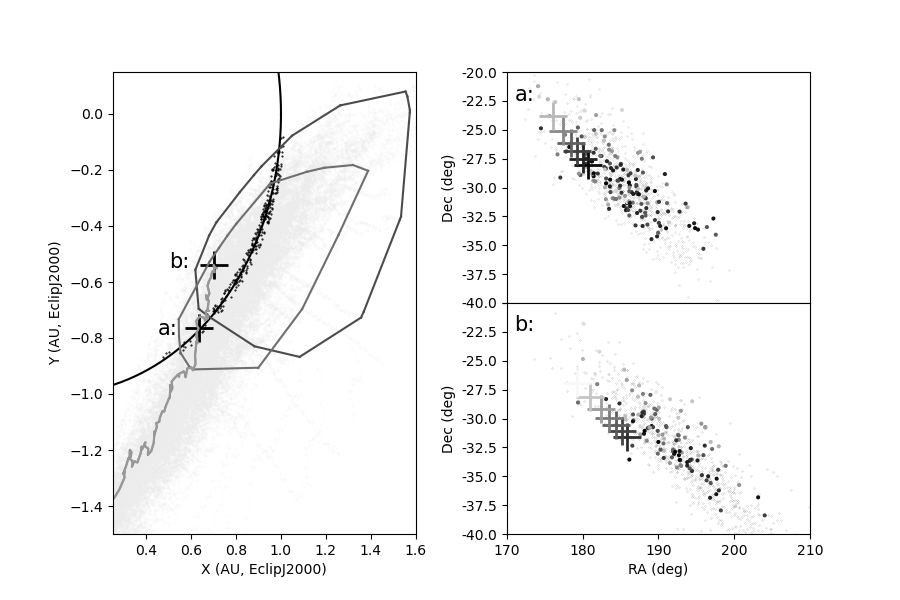}
    \caption{Locations of meteoroid and comet node crossings in the ecliptic plane (left) and the projected location of comet and meteoroids on the sky at two epochs (right, see text for details) for a simulation of shower $\#206$ (the Aurigids). The parent body is marked in both panels with crosses.  The location of approximately 95\% of meteors crossing within 500 years of the comet at epochs 'a' and 'b' are encircled (left), or appear in light gray (right).  The subset of Earth-intersecting ejecta are shown in black on the left, and grayscale graded by magnitude (right).  On the right, the comet's location is shown at Vernal Equinox for all the years it takes to traverse from $V=25$ to $V=22$.}
    \label{fig_validation_1}
\end{figure*}

\section{Discussion} 
\label{sec:discussion}

\subsection{Uncertainty of the search area} 
\label{sec:4.1}

While we found that the calculated search areas capture well the location of the comet for showers with known parent bodies, there are some matters that can change the search area in other cases. We already mentioned that our method only considers the dispersion of the meteoroid stream in Earth's path, not the full dispersion of the stream. The methodology assumes that the orbital elements of a meteoroid stream are similar to those of the parent body itself. It also assumes that the meteor shower elements are independently distributed, each characterized by only a median and dispersion, and that Keplerian orbits provide a good approximation when backing up the particles from their node to some previous time.  These assumptions would be fraught for short period comets ($P<200$ yr), where planetary encounters and resonances might be expected to alter the parent body's orbit significantly from the shower on short (decadal) timescales, but they seem reasonable for LPCs.

Even for long-period comets, the actual distributions of orbital elements in the meteor stream are not Gaussian, and likely correlated, so that independent Gaussian deviates may not adequately represent the location of the cloud on the sky. This comes into play mainly for low-inclined orbits that precess at a fast rate.

Finally, the dispersion of orbital elements may not adequately represent the measured dispersion of orbits defined by the radiant and velocity of meteors in a shower \citep{ms_lpc}. We conducted simulations starting from radiant and speed distributions to see if this would provide any benefit in terms of generating smaller clouds, and therefore smaller search regions on sky. We found that both approaches gave similar results: 57\% of these radiant-derived particles were distributed in somewhat smaller clouds and 43\% resulted in larger clouds. The clouds from these radiant-derived elements were at best $\sim$50\% smaller and at worst 300\% larger than orbital element generated clouds, but overall there was not a systematic difference. Because of that, we present results here using the median Keplerian orbital elements and their dispersions as shown in Table \ref{table1}.

\subsection{Separate evolution of comet and its meteoroid stream} 
\label{sec:4.2}

Orbits will be perturbed from the Keplerian orbits during perihelion crossing and, over long periods of time, this can create a stream in Earth's path that does not fully describe the extent of the stream away from Earth's orbit. The method failed for shower \#176, if the proposed parent comet is indeed the parent body of that stream. 

We performed numerical simulations of LPC dynamical evolution to investigate if the comet orbit stays among the meteoroid orbits over the age of the observed streams \citep{Pilorz_2023}. Examination of the loci of nodal crossings for the parent body and ejecta cloud shows that the comet has a node within the cloud's nodes, though both undergo nodal progression and stochastic wandering.  

Figure~\ref{fig_validation_1} shows results from simulations of Shower $\#206$ (the Aurigids), at two different epochs.  For those simulations, a model for the parent comet C/1911 N1 (Kiess) was created for times between 60kyr ago and the present, the timescale on which LPC ejecta are thought to disperse. It was found by performing backward followed by forward integration of several perturbations of the current observed elements, from which the variant was selected whose forward integration resulted in present-day orbital elements most similar to the observed, allowing a period between 250 and 4000 years.  Numerical integrations were then performed of ejecta released during selected perihelia of this model comet, following the general method of \cite{Vaubaillon_2005}.  In this case, 300 ejecta were released uniformly in time when the comet was within 3 au of the Sun, with velocities uniformly within $1-10\,{\rm m s}^{-1}$ and random directions over the lit face.

The left-hand panel of Figure~\ref{fig_validation_1} shows the loci of all node crossings of the cloud of 300 particles in ecliptic J2000 coordinates, along with the parent body at two epochs marked 'a:' and 'b:'. The trajectory of the comet's node location is drawn as a solid curve, and effects of nodal progression and stochastic wandering are apparent. The crosses show the locations of the node at an epoch near Earth crossing, and a later epoch at which it is inside Earth's orbit.  Polygons enclose approximately 95\% of the ejecta that crossed the node within 500 years of the comet's crossing at each epoch.  The subset of these that intersect Earth's orbit, shown in black dots, are taken to be what an observer at that epoch would use to model the stream orbital elements.  

The right-hand panels of Figure~\ref{fig_validation_1} show the result of numerically integrating that subset of points backwards individually, to locations on their inbound orbits at which they would have apparent magnitudes within $22 < V < 25$, with the end magnitude for each point chosen randomly within that range. Those magnitudes correspond to ranges between approximately 12-25 au, and take the comet approximately seven years to traverse. The seven crosses are the comet's RA/Dec at vernal equinox each year for those seven years.  The crosses and points are shaded by their apparent magnitudes, with black corresponding to $V=25,$ and lighter grey to $V=22.$  The particles' locations move from the lower right to upper left of the cloud as they approach.  

This figure shows typical behavior we observe: the comet's location on the sky lies within the cloud of points at epochs where its node is near 1 au, but lies at the edge of the cloud at epochs when the node has drifted inwards. 

The numerical simulations indicate that even for meteors ejected 60 kyr in the past, the position of the comet remains within or at the edge of the stream. Its location on the sky is within the cloud of the subset of ejecta that cross Earth's orbit and would correspond to the sampled distributions described in Section~\ref{sec:2.3}

\begin{deluxetable}{ccccccc}[bp]
    \tablecaption{Orbital elements and standard deviations of meteor shower $\#16$ ($\sigma$-Hydrids), from \citet{Jenniskens2023}. \label{hydrids}}
    \savetablenum{4}
    \tablehead{\colhead{} & \colhead{} & \colhead{} & \colhead{Value} & \colhead{} & \colhead{} & \colhead{Dispersion}}
    \startdata
        q & & & 0.257 au & & & \textpm0.0437 au \\
        e & & & 0.986 & & & \textpm0.0596 \\
        i & & & 128.8° & & & \textpm3.72° \\
        $\omega$ & & & 119.3° & & & \textpm7.15° \\
        $\Omega$ & & & 76.6° & & & \textpm8.30° \\
    \enddata
\end{deluxetable}

\subsection{Comet Nishimura} 
\label{sec:4.3}

C/2023 P1 (Nishimura) was a long period comet discovered by amateur astronomer Hideo Nishimura with a digital camera in August 2023 \citep{mpec_nishimura}. The comet was very low on the horizon and escaped automatic detection by sky-surveys. In combing through sky-survey data after discovery of Comet Nishimura, serendipitous observations were found in Pan-STARRS data dating back to January 2023 \citep{mpc_nishimura}. Comet Nishimura is also a proposed parent body of the $\sigma$-Hydrids meteor shower \citep{CBET_ye}. This presented an opportunity to test our modelling on direct observations of a newly discovered long period comet with an observed meteor shower. 

There are no published estimates of Comet Nishimura's absolute magnitude, so we ran simulations across the range of $H$ used for the previous simulations ($10 \le H \le 17$). Using the orbital elements of the $\sigma$-Hydrids, their measured dispersions (Table \ref{hydrids}) and an assumed $H$, synthetic comets were created and ephemerides were calculated with \fontfamily{cmtt}\selectfont OpenOrb \fontfamily{cmr}\selectfont using obscode F52 (Pan-STARRS 2, Haleakala) for the date of the earliest Pan-STARRS detection (2023-01-19.326034). From the JPL Horizons Small Body database, the comet was determined to be 4.0 au from the sun at that time. This provided a basis to bound the heliocentric distance of synthetic comets and set observability to be those with $3 \le r \le 5$ (1 au on either side gives a smaller cloud of potential positions) and on their inbound trajectory.

\begin{figure}
    \centering
    \includegraphics[width=0.5\textwidth]{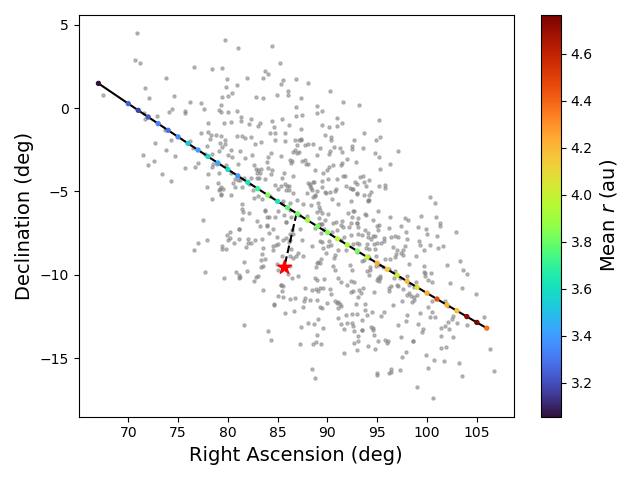}
    \caption{Search area on sky for shower $\#16$ ($\sigma$-Hydrids) on 2023-01-19.326034 (UT). The synthetic comets in this simulation have an assumed $H=16$. Its parent comet, C/2023 P1 (Nishimura) is represented by the red star. The model results show $\delta_l$ = 3.43°, $\Psi$ = 8.25°, $\psi_l$ = 42\%, and $\delta_r$ = 0.24 au.}
    \label{hydrids_search}
\end{figure}

Figure \ref{hydrids_search} provides a visualization of the search region on sky produced by the $\sigma$-Hydrids meteor shower with an assumed $H=16$. Comet Nishimura's position on 2023-01-19.326034 (UT) was pulled from the Pan-STARRS observation. We found $\delta_l$ to be 3.43°, $\Psi= 8.25$°, and $\delta_r$ was 0.24 au. The comet fell at a $\psi_l$ of 42\%. We obtained a range for $\delta_l$ = 2.98° - 3.59° across all $H$ values. 

\begin{figure}
    \centering
    \includegraphics[width=0.5\textwidth]{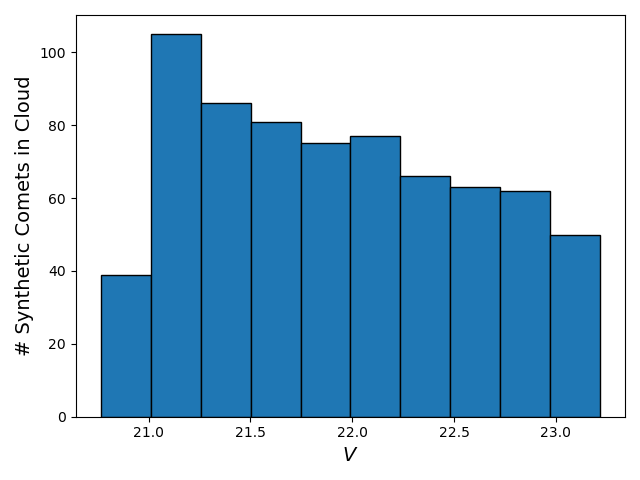}
    \caption{Histogram of visual magnitudes of synthetic comets in the search region produced by shower $\#16$ ($\sigma$-Hydrids) as seen in Figure \ref{hydrids_search}.}
    \label{hydrids_hist}
\end{figure}

Figure \ref{hydrids_hist} shows a distribution of the visual magnitudes of observable synthetic comets from the modelling with $H=16$. Reported G-band magnitudes from pre-covery images on 2023-01-19 were between 21.24 and 21.64 in 4 observations \citep{mpc_nishimura}. The distribution of $V$ for each simulation was where we saw the largest variation due to $H$. $H=10$ produced synthetic comets with $14.8 \le V \le 17.2$ whereas $H=17$ produced $21.8 \le V \le 24.2$. An $H\approx16$ appears to be the best fit for this comet, using measurements from the observations to be a basis for the expected brightness of synthetic comets. 

Our model assumes that the comet was inactive in the Pan-STARRS observations. Pre-covery images from Pan-STARRS showed that the comet was not displaying signs of activity \citep{CBET_weryk}. Using Nishimura's $r$ (4.02 au), $\Delta$ (3.27 au), and $\alpha$ (10.0°) on the observation date, equation \ref{eqn1} returns an apparent magnitude of 21.5 for an inactive comet nucleus with $H=16$. This calculated $V$ from the assumed $H$ is consistent with our nominal model and direct observation.

In this exercise, known information about Nishimura helped guide which synthetic comets were observable. This highlights where the process would begin when attempting to set constraints on a blind search. When such a search for a comet is being conducted, an observer would start by knowing information about the limiting magnitude of their instrument. An estimate of the comet's absolute magnitude would need to be taken into consideration to guide the heliocentric distance range of interest for synthetic comets in the model. This could be done blindly, by using a distribution as done in this work, or possibly by using meteor shower activity to derive an estimate of $H$. 

An instrument with a wide field of view would be in the observer's best interest. With a simulated full cloud width approaching 20° for the $\sigma$-Hydrids, an instrument with a 3° FOV would be able to capture the full width of the search region in $\sim$7 images. Exposure times for each image would factor in to how long it might take to conduct a search in any particular region, thereby allowing observers to assess the feasibility of such a search. This method can provide a starting point to select sky-survey observations of interesting search regions.

\subsection{The comet association with a meteor shower} 
\label{sec:4.4}

Not all comets in our sample may be correctly identified as the parent bodies of given meteor showers. C/2015 D4, with an uncertain association to shower $\#176$, showed the furthest average line distance of all showers in our analysis by $\sim$10°. However, with an $H$ of 14.8 listed in the JPL Small Body database, the number of simulations run in the analysis for this shower was small and not enough to definitively determine whether this comet is in fact related to the meteor shower.

In a real search for parent comets, it would be likely that many objects could be detected in these large search regions. We note that the mere detection of an object in one of these regions does not imply that it is the parent body of a meteor shower. These detections will identify objects that are candidates for follow-up observations. As the orbit and trajectory of the object are further refined, D-criterion tests can be computed to confirm or deny a parent/shower relationship.

Determining the number of objects that might ‘look’ like the objects for which we are searching can give us insight into the probability of a false association. We used the orbits of 831 comets with ‘C’ prefixes and eccentricities $\le$ 1 from the JPL Small Body Database to quantify the possible rate of false detections. We determined which orbits intersected the simulated clouds in Figure \ref{celestial_sphere}. From there, we looked at which of these comets were within the brightness constraints we set for our searches $(22 \leq V \leq 25)$ and moving within the velocity constraints found in our analysis. These constraints are non-sidereal rates \textpm0.15 \arcsec/min and position angle (east of north) \textpm15° of the synthetic comet’s mean motion within a 3° box around the comet as it is moved through the cloud of synthetic objects.

When applying these position and velocity cuts, we found an average of 11.4 of the 831 comets ($1.4\%$) could be mistaken for the parent comet across all 17 showers. False probabilities were as high as $4.1\%$ for shower \#176 and as low as $0.2\%$ for showers \#545 and \#541. The known population skews the size distribution towards larger objects $(H \leq 17)$ and may not fully reflect the total comet population. Given that, this exercise shows how a large number of objects $(\geq 95\%)$ detected in these regions may be ruled out based on the expected brightness and direction of motion calculated by the simulated shower.

\section{Conclusions} 
\label{sec:conclusions}

Meteor showers can be a practical guide to dedicated long-period comet searches for long-period comets with the range of orbital periods 200 to 4000 years. By identifying comets at an earlier stage of their approach towards Earth, this would increase the warning time to help devise and implement more effective strategies for deflecting an impact. 

Our dynamical modeling shows that the parent comet and meteoroids evolve in much the same way and that in most cases the parent comet is expected to be found in projection among the synthetic comets sampled from the observed meteor shower in Earth's path. 

Indeed, known parent comets of meteor showers are mostly found inside those search areas. On average, the parent comets would have been discovered at a distance of 1.51° $\pm$1.19° from the mean positions of a cloud of synthetic comets. 50\% of the comets fell at $\delta_l = 1.18$° ($\psi_l = 30.3\%$), 70\% fell at $\delta_l = 1.91$° ($\psi_l = 47.9\%$), and 90\% fell at $\delta_l = 3.29$° ($\psi_l = 84.5\%$).

The uncertain brightness of the potential parent body makes the search areas elongated along the projected comet orbit on the sky. The mean heliocentric distance of synthetic comet particles from the trend line was 2.09 $\pm1.89$ au for showers with unknown $H$'s and 0.96 $\pm0.80$ au for showers with known $H$'s. 

The search areas calculated from the meteoroid stream dispersions at Earth are feasible for dedicated searches. The $\delta_l$'s calculated in this study confirm the feasibility of using wide angle sky surveys, such as LSST (3.5° FOV) or DECam (2.2° FOV), to discover unknown parent comets of long period comet meteor showers. 

The meteor showers also constrain the motion of the parent comet on sky. Within a 3° region, the parent comets were travelling within \textpm0.15 \arcsec/min of the mean rate and \textpm15° of the mean position angle of synthetic comets in that region. With typical seeing conditions of $\sim$1$\arcsec$ and comets travelling 0.2 \arcsec/min or less, the feasibility of discovery is not likely to be influenced by trailing losses. These values indicate that exposures of up to 5 minutes can be taken before trailing losses would effect the discovery of these objects.

Detecting LPCs at their faintest can provide years worth of warning time for a potential impactor. The warning time can increase to over 10 years by detecting the largest objects with this method. For smaller objects, the additional year or two gained by earlier detection would still give significantly more time to mitigate an impact. 

This method successfully used the $\sigma$-Hydrids to create a search region matching the location of Comet Nishimura in pre-covery images taken 8 months prior to its discovery.

\subsection{Future work} 
\label{sec:5.1}

This method can now be applied to LSST observations. Expanding our knowledge and catalog of small bodies in the solar system is a main science driver of LSST \citep{Ivezic_2019}. It is predicted that LSST will more than double the current comet population from $\sim$4000 to $\sim$10,000 objects \citep{science_impact_LSST_SSscience}. LSST simulations can be run to determine how many parents of these long period comet meteor showers may be discoverable with this survey. This information could then be used to prioritize specific search regions within the LSST data stream.  

The ideal search strategy would be to push the detection limits of LSST by employing a shift-and-stack search method. The meteor shower informs which region of the sky the parent will be in as well as its speed and direction of motion. Warning time in the case of a PHO is also improved by pushing the limits of detection beyond that of a single image exposure.

The LSST data stream will generate $\sim$400 alerts per visit associated with moving objects\footnote[1]{https://dmtn-102.lsst.io/DMTN-102.pdf}. Alerts are distributed within 60 seconds of camera readout to designated alert brokers, which are then accessible by the scientific community\footnote[2]{https://ldm-612.lsst.io/LDM-612.pdf}. With ~1000 visits per night, the sheer volume of alerts generated in a single night will be overwhelming. In addition to the number of alerts generated by these detections, faint discoveries barely above background noise in LSST images will likely escape automatic detection algorithms. 

Knowing which regions of sky will contain objects of interest will help prioritize analysis steps. Our method will guide searches for objects in individual exposures that were missed in automated detection routines and can be used to probe deeper with image stacking techniques that will not be part of the standard LSST data processing pipeline.

As a first step, \cite{Jenniskens2023} determined the orbital elements and dispersions of 247 long-period comet meteoroid streams, most from unknown long period comet parents. The development of a website to guide searches for these objects is in the early development phases.

As a future improvement on the method outlined here, we aim to determine probability densities for discovery within each shower's search region. With an average solid angle of 255 steradians per region, this can highlight specific areas within the search regions that would have higher probability for success. One method to assign probability would be to weight the synthetic comets by the inverse of their orbital period. Longer orbit synthetic comets may better resemble the parent comet's orbit. This is because the measured orbits of meteor showers are often shorter than their parent comets due to more frequent Earth impacts. This refinement could further narrow down the best areas to begin a search.

\begin{acknowledgments}
\section{Acknowledgments}
The authors would like to thank two anonymous reviewers for their constructive feedback which enhanced the quality of this work. PJ and SP acknowledge support from NASA YORPD grant 80NSSC22K1467. SH and NM acknowledge support from NSF grant $\#AST1944827$ through the CAREER program (PI N. Moskovitz) and the Marcus Comet Fund. 
\end{acknowledgments}

\software{\fontfamily{cmtt}\selectfont Astropy \fontfamily{cmr}\selectfont \citep{astropy:2013, astropy:2018, astropy:2022}, \fontfamily{cmtt}\selectfont
astroquery \fontfamily{cmr}\selectfont \citep{Ginsburg_2019} \fontfamily{cmtt}\selectfont Matplotlib \fontfamily{cmr}\selectfont \citep{Hunter:2007}, \fontfamily{cmtt}\selectfont Numpy \fontfamily{cmr}\selectfont \citep{harris2020array}, \fontfamily{cmtt}\selectfont
OpenOrb \fontfamily{cmr}\selectfont \citep{OpenOrb}}




\bibliography{refs}{}

\begin{thebibliography}{}
\expandafter\ifx\csname natexlab\endcsname\relax\def\natexlab#1{#1}\fi
\providecommand{\url}[1]{\href{#1}{#1}}
\providecommand{\dodoi}[1]{doi:~\href{http://doi.org/#1}{\nolinkurl{#1}}}
\providecommand{\doeprint}[1]{\href{http://ascl.net/#1}{\nolinkurl{http://ascl.net/#1}}}
\providecommand{\doarXiv}[1]{\href{https://arxiv.org/abs/#1}{\nolinkurl{https://arxiv.org/abs/#1}}}

\bibitem[{{Astropy Collaboration} {et~al.}(2013){Astropy Collaboration},
  {Robitaille}, {Tollerud}, {Greenfield}, {Droettboom}, {Bray}, {Aldcroft},
  {Davis}, {Ginsburg}, {Price-Whelan}, {Kerzendorf}, {Conley}, {Crighton},
  {Barbary}, {Muna}, {Ferguson}, {Grollier}, {Parikh}, {Nair}, {Unther},
  {Deil}, {Woillez}, {Conseil}, {Kramer}, {Turner}, {Singer}, {Fox}, {Weaver},
  {Zabalza}, {Edwards}, {Azalee Bostroem}, {Burke}, {Casey}, {Crawford},
  {Dencheva}, {Ely}, {Jenness}, {Labrie}, {Lim}, {Pierfederici}, {Pontzen},
  {Ptak}, {Refsdal}, {Servillat}, \& {Streicher}}]{astropy:2013}
{Astropy Collaboration}, {Robitaille}, T.~P., {Tollerud}, E.~J., {et~al.} 2013,
  \aap, 558, A33, \dodoi{10.1051/0004-6361/201322068}

\bibitem[{{Astropy Collaboration} {et~al.}(2018){Astropy Collaboration},
  {Price-Whelan}, {Sip{\H{o}}cz}, {G{\"u}nther}, {Lim}, {Crawford}, {Conseil},
  {Shupe}, {Craig}, {Dencheva}, {Ginsburg}, {Vand erPlas}, {Bradley},
  {P{\'e}rez-Su{\'a}rez}, {de Val-Borro}, {Aldcroft}, {Cruz}, {Robitaille},
  {Tollerud}, {Ardelean}, {Babej}, {Bach}, {Bachetti}, {Bakanov}, {Bamford},
  {Barentsen}, {Barmby}, {Baumbach}, {Berry}, {Biscani}, {Boquien}, {Bostroem},
  {Bouma}, {Brammer}, {Bray}, {Breytenbach}, {Buddelmeijer}, {Burke},
  {Calderone}, {Cano Rodr{\'\i}guez}, {Cara}, {Cardoso}, {Cheedella}, {Copin},
  {Corrales}, {Crichton}, {D'Avella}, {Deil}, {Depagne}, {Dietrich}, {Donath},
  {Droettboom}, {Earl}, {Erben}, {Fabbro}, {Ferreira}, {Finethy}, {Fox},
  {Garrison}, {Gibbons}, {Goldstein}, {Gommers}, {Greco}, {Greenfield},
  {Groener}, {Grollier}, {Hagen}, {Hirst}, {Homeier}, {Horton}, {Hosseinzadeh},
  {Hu}, {Hunkeler}, {Ivezi{\'c}}, {Jain}, {Jenness}, {Kanarek}, {Kendrew},
  {Kern}, {Kerzendorf}, {Khvalko}, {King}, {Kirkby}, {Kulkarni}, {Kumar},
  {Lee}, {Lenz}, {Littlefair}, {Ma}, {Macleod}, {Mastropietro}, {McCully},
  {Montagnac}, {Morris}, {Mueller}, {Mumford}, {Muna}, {Murphy}, {Nelson},
  {Nguyen}, {Ninan}, {N{\"o}the}, {Ogaz}, {Oh}, {Parejko}, {Parley}, {Pascual},
  {Patil}, {Patil}, {Plunkett}, {Prochaska}, {Rastogi}, {Reddy Janga},
  {Sabater}, {Sakurikar}, {Seifert}, {Sherbert}, {Sherwood-Taylor}, {Shih},
  {Sick}, {Silbiger}, {Singanamalla}, {Singer}, {Sladen}, {Sooley},
  {Sornarajah}, {Streicher}, {Teuben}, {Thomas}, {Tremblay}, {Turner},
  {Terr{\'o}n}, {van Kerkwijk}, {de la Vega}, {Watkins}, {Weaver}, {Whitmore},
  {Woillez}, {Zabalza}, \& {Astropy Contributors}}]{astropy:2018}
{Astropy Collaboration}, {Price-Whelan}, A.~M., {Sip{\H{o}}cz}, B.~M., {et~al.}
  2018, \aj, 156, 123, \dodoi{10.3847/1538-3881/aabc4f}

\bibitem[{{Astropy Collaboration} {et~al.}(2022){Astropy Collaboration},
  {Price-Whelan}, {Lim}, {Earl}, {Starkman}, {Bradley}, {Shupe}, {Patil},
  {Corrales}, {Brasseur}, {N{"o}the}, {Donath}, {Tollerud}, {Morris},
  {Ginsburg}, {Vaher}, {Weaver}, {Tocknell}, {Jamieson}, {van Kerkwijk},
  {Robitaille}, {Merry}, {Bachetti}, {G{"u}nther}, {Aldcroft},
  {Alvarado-Montes}, {Archibald}, {B{'o}di}, {Bapat}, {Barentsen}, {Baz{'a}n},
  {Biswas}, {Boquien}, {Burke}, {Cara}, {Cara}, {Conroy}, {Conseil}, {Craig},
  {Cross}, {Cruz}, {D'Eugenio}, {Dencheva}, {Devillepoix}, {Dietrich},
  {Eigenbrot}, {Erben}, {Ferreira}, {Foreman-Mackey}, {Fox}, {Freij}, {Garg},
  {Geda}, {Glattly}, {Gondhalekar}, {Gordon}, {Grant}, {Greenfield}, {Groener},
  {Guest}, {Gurovich}, {Handberg}, {Hart}, {Hatfield-Dodds}, {Homeier},
  {Hosseinzadeh}, {Jenness}, {Jones}, {Joseph}, {Kalmbach}, {Karamehmetoglu},
  {Ka{l}uszy{'n}ski}, {Kelley}, {Kern}, {Kerzendorf}, {Koch}, {Kulumani},
  {Lee}, {Ly}, {Ma}, {MacBride}, {Maljaars}, {Muna}, {Murphy}, {Norman},
  {O'Steen}, {Oman}, {Pacifici}, {Pascual}, {Pascual-Granado}, {Patil},
  {Perren}, {Pickering}, {Rastogi}, {Roulston}, {Ryan}, {Rykoff}, {Sabater},
  {Sakurikar}, {Salgado}, {Sanghi}, {Saunders}, {Savchenko}, {Schwardt},
  {Seifert-Eckert}, {Shih}, {Jain}, {Shukla}, {Sick}, {Simpson},
  {Singanamalla}, {Singer}, {Singhal}, {Sinha}, {Sip{H{o}}cz}, {Spitler},
  {Stansby}, {Streicher}, {{{S}}umak}, {Swinbank}, {Taranu}, {Tewary},
  {Tremblay}, {Val-Borro}, {Van Kooten}, {Vasovi{'c}}, {Verma}, {de Miranda
  Cardoso}, {Williams}, {Wilson}, {Winkel}, {Wood-Vasey}, {Xue}, {Yoachim},
  {Zhang}, {Zonca}, \& {Astropy Project Contributors}}]{astropy:2022}
{Astropy Collaboration}, {Price-Whelan}, A.~M., {Lim}, P.~L., {et~al.} 2022,
  apj, 935, 167, \dodoi{10.3847/1538-4357/ac7c74}

\bibitem[{Bellm {et~al.}(2018)Bellm, Kulkarni, Graham, Dekany, Smith, Riddle,
  Masci, Helou, Prince, Adams, Barbarino, Barlow, Bauer, Beck, Belicki, Biswas,
  Blagorodnova, Bodewits, Bolin, Brinnel, Brooke, Bue, Bulla, Burruss, Cenko,
  Chang, Connolly, Coughlin, Cromer, Cunningham, De, Delacroix, Desai, Duev,
  Eadie, Farnham, Feeney, Feindt, Flynn, Franckowiak, Frederick, Fremling,
  Gal-Yam, Gezari, Giomi, Goldstein, Golkhou, Goobar, Groom, Hacopians, Hale,
  Henning, Ho, Hover, Howell, Hung, Huppenkothen, Imel, Ip, {\v{Z}eljko
  Ivezi{\'c}}, Jackson, Jones, Juric, Kasliwal, Kaspi, Kaye, Kelley, Kowalski,
  Kramer, Kupfer, Landry, Laher, Lee, Lin, Lin, Lunnan, Giomi, Mahabal, Mao,
  Miller, Monkewitz, Murphy, Ngeow, Nordin, Nugent, Ofek, Patterson, Penprase,
  Porter, Rauch, Rebbapragada, Reiley, Rigault, Rodriguez, van Roestel,
  Rusholme, van Santen, Schulze, Shupe, Singer, Soumagnac, Stein, Surace,
  Sollerman, Szkody, Taddia, Terek, Sistine, van Velzen, Vestrand, Walters,
  Ward, Ye, Yu, Yan, \& Zolkower}]{Bellm_2019}
Bellm, E.~C., Kulkarni, S.~R., Graham, M.~J., {et~al.} 2018, Publications of
  the Astronomical Society of the Pacific, 131, 018002,
  \dodoi{10.1088/1538-3873/aaecbe}

\bibitem[{Bernstein {et~al.}(2004)Bernstein, Trilling, Allen, Brown, Holman, \&
  Malhotra}]{Bernstein_2004}
Bernstein, G.~M., Trilling, D.~E., Allen, R.~L., {et~al.} 2004, The
  Astronomical Journal, 128, 1364, \dodoi{10.1086/422919}

\bibitem[{Boe {et~al.}(2019)Boe, Jedicke, Meech, Wiegert, Weryk, Chambers,
  Denneau, Kaiser, Kudritzki, Magnier, Wainscoat, \& Waters}]{BOE2019252}
Boe, B., Jedicke, R., Meech, K.~J., {et~al.} 2019, Icarus, 333, 252,
  \dodoi{https://doi.org/10.1016/j.icarus.2019.05.034}

\bibitem[{Chambers {et~al.}(2019)Chambers, Magnier, Metcalfe, Flewelling,
  Huber, Waters, Denneau, Draper, Farrow, Finkbeiner, Holmberg, Koppenhoefer,
  Price, Rest, Saglia, Schlafly, Smartt, Sweeney, Wainscoat, Burgett, Chastel,
  Grav, Heasley, Hodapp, Jedicke, Kaiser, Kudritzki, Luppino, Lupton, Monet,
  Morgan, Onaka, Shiao, Stubbs, Tonry, White, Bañados, Bell, Bender, Bernard,
  Boegner, Boffi, Botticella, Calamida, Casertano, Chen, Chen, Cole, Deacon,
  Frenk, Fitzsimmons, Gezari, Gibbs, Goessl, Goggia, Gourgue, Goldman, Grant,
  Grebel, Hambly, Hasinger, Heavens, Heckman, Henderson, Henning, Holman, Hopp,
  Ip, Isani, Jackson, Keyes, Koekemoer, Kotak, Le, Liska, Long, Lucey, Liu,
  Martin, Masci, McLean, Mindel, Misra, Morganson, Murphy, Obaika, Narayan,
  Nieto-Santisteban, Norberg, Peacock, Pier, Postman, Primak, Rae, Rai, Riess,
  Riffeser, Rix, Röser, Russel, Rutz, Schilbach, Schultz, Scolnic, Strolger,
  Szalay, Seitz, Small, Smith, Soderblom, Taylor, Thomson, Taylor, Thakar,
  Thiel, Thilker, Unger, Urata, Valenti, Wagner, Walder, Walter, Watters,
  Werner, Wood-Vasey, \& Wyse}]{chambers2019panstarrs1}
Chambers, K.~C., Magnier, E.~A., Metcalfe, N., {et~al.} 2019, The Pan-STARRS1
  Surveys.
\newblock \doarXiv{1612.05560}

\bibitem[{Drummond(1981)}]{DRUMMOND1981545}
Drummond, J.~D. 1981, Icarus, 45, 545,
  \dodoi{https://doi.org/10.1016/0019-1035(81)90020-8}

\bibitem[{Flaugher {et~al.}(2015)Flaugher, Diehl, Honscheid, Abbott, Alvarez,
  Angstadt, Annis, Antonik, Ballester, Beaufore, Bernstein, Bernstein, Bigelow,
  Bonati, Boprie, Brooks, Buckley-Geer, Campa, Cardiel-Sas, Castander,
  Castilla, Cease, Cela-Ruiz, Chappa, Chi, Cooper, da~Costa, Dede, Derylo,
  DePoy, de~Vicente, Doel, Drlica-Wagner, Eiting, Elliott, Emes, Estrada, Neto,
  Finley, Flores, Frieman, Gerdes, Gladders, Gregory, Gutierrez, Hao, Holland,
  Holm, Huffman, Jackson, James, Jonas, Karcher, Karliner, Kent, Kessler,
  Kozlovsky, Kron, Kubik, Kuehn, Kuhlmann, Kuk, Lahav, Lathrop, Lee, Levi,
  Lewis, Li, Mandrichenko, Marshall, Martinez, Merritt, Miquel, Muñoz,
  Neilsen, Nichol, Nord, Ogando, Olsen, Palaio, Patton, Peoples, Plazas, Rauch,
  Reil, Rheault, Roe, Rogers, Roodman, Sanchez, Scarpine, Schindler, Schmidt,
  Schmitt, Schubnell, Schultz, Schurter, Scott, Serrano, Shaw, Smith,
  Soares-Santos, Stefanik, Stuermer, Suchyta, Sypniewski, Tarle, Thaler, Tighe,
  Tran, Tucker, Walker, Wang, Watson, Weaverdyck, Wester, Woods, Yanny, \&
  Collaboration)}]{Flaugher_2015}
Flaugher, B., Diehl, H.~T., Honscheid, K., {et~al.} 2015, The Astronomical
  Journal, 150, 150, \dodoi{10.1088/0004-6256/150/5/150}

\bibitem[{Ginsburg {et~al.}(2019)Ginsburg, Sipőcz, Brasseur, Cowperthwaite,
  Craig, Deil, Groener, Guillochon, Guzman, Liedtke, Lim, Lockhart, Mommert,
  Morris, Norman, Parikh, Persson, Robitaille, Segovia, Singer, Tollerud,
  de~Val-Borro, Valtchanov, Woillez, \& The
  Astroquery~collaboration}]{Ginsburg_2019}
Ginsburg, A., Sipőcz, B.~M., Brasseur, C.~E., {et~al.} 2019, The Astronomical
  Journal, 157, 98, \dodoi{10.3847/1538-3881/aafc33}

\bibitem[{Granvik {et~al.}(2009)Granvik, Virtanen, Oszkiewicz, \&
  Muinonen}]{OpenOrb}
Granvik, M., Virtanen, J., Oszkiewicz, D., \& Muinonen, K. 2009, Meteoritics \&
  Planetary Science, 44, 1853

\bibitem[{Gritzner {et~al.}(2006)Gritzner, D{\"u}rfeld, Kasper, \&
  Fasoulas}]{Gritzner_2006}
Gritzner, C., D{\"u}rfeld, K., Kasper, J., \& Fasoulas, S. 2006,
  Naturwissenschaften, 93, 361, \dodoi{10.1007/s00114-006-0115-0}

\bibitem[{Groussin {et~al.}(2010)Groussin, Lamy, \& Jorda}]{GROUSSIN2010904}
Groussin, O., Lamy, P., \& Jorda, L. 2010, Planetary and Space Science, 58,
  904, \dodoi{https://doi.org/10.1016/j.pss.2010.02.009}

\bibitem[{Harris {et~al.}(2020)Harris, Millman, van~der Walt, Gommers,
  Virtanen, Cournapeau, Wieser, Taylor, Berg, Smith, Kern, Picus, Hoyer, van
  Kerkwijk, Brett, Haldane, del R{\'{i}}o, Wiebe, Peterson,
  G{\'{e}}rard-Marchant, Sheppard, Reddy, Weckesser, Abbasi, Gohlke, \&
  Oliphant}]{harris2020array}
Harris, C.~R., Millman, K.~J., van~der Walt, S.~J., {et~al.} 2020, Nature, 585,
  357, \dodoi{10.1038/s41586-020-2649-2}

\bibitem[{Hunter(2007)}]{Hunter:2007}
Hunter, J.~D. 2007, Computing in Science \& Engineering, 9, 90,
  \dodoi{10.1109/MCSE.2007.55}

\bibitem[{Ivezi{\'c} {et~al.}(2019)Ivezi{\'c}, Kahn, Tyson, Abel, Acosta,
  Allsman, Alonso, AlSayyad, Anderson, Andrew, Angel, Angeli, Ansari,
  Antilogus, Araujo, Armstrong, Arndt, Astier, Éric Aubourg, Auza, Axelrod,
  Bard, Barr, Barrau, Bartlett, Bauer, Bauman, Baumont, Bechtol, Bechtol,
  Becker, Becla, Beldica, Bellavia, Bianco, Biswas, Blanc, Blazek, Blandford,
  Bloom, Bogart, Bond, Booth, Borgland, Borne, Bosch, Boutigny, Brackett,
  Bradshaw, Brandt, Brown, Bullock, Burchat, Burke, Cagnoli, Calabrese,
  Callahan, Callen, Carlin, Carlson, Chandrasekharan, Charles-Emerson, Chesley,
  Cheu, Chiang, Chiang, Chirino, Chow, Ciardi, Claver, Cohen-Tanugi, Cockrum,
  Coles, Connolly, Cook, Cooray, Covey, Cribbs, Cui, Cutri, Daly, Daniel,
  Daruich, Daubard, Daues, Dawson, Delgado, Dellapenna, de~Peyster,
  de~Val-Borro, Digel, Doherty, Dubois, Dubois-Felsmann, Durech, Economou,
  Eifler, Eracleous, Emmons, Neto, Ferguson, Figueroa, Fisher-Levine, Focke,
  Foss, Frank, Freemon, Gangler, Gawiser, Geary, Gee, Geha, Gessner, Gibson,
  Gilmore, Glanzman, Glick, Goldina, Goldstein, Goodenow, Graham, Gressler,
  Gris, Guy, Guyonnet, Haller, Harris, Hascall, Haupt, Hernandez, Herrmann,
  Hileman, Hoblitt, Hodgson, Hogan, Howard, Huang, Huffer, Ingraham, Innes,
  Jacoby, Jain, Jammes, Jee, Jenness, Jernigan, Jevremović, Johns, Johnson,
  Johnson, Jones, Juramy-Gilles, Jurić, Kalirai, Kallivayalil, Kalmbach,
  Kantor, Karst, Kasliwal, Kelly, Kessler, Kinnison, Kirkby, Knox, Kotov,
  Krabbendam, Krughoff, Kubánek, Kuczewski, Kulkarni, Ku, Kurita, Lage,
  Lambert, Lange, Langton, Guillou, Levine, Liang, Lim, Lintott, Long, Lopez,
  Lotz, Lupton, Lust, MacArthur, Mahabal, Mandelbaum, Markiewicz, Marsh,
  Marshall, Marshall, May, McKercher, McQueen, Meyers, Migliore, Miller, Mills,
  Miraval, Moeyens, Moolekamp, Monet, Moniez, Monkewitz, Montgomery, Morrison,
  Mueller, Muller, Arancibia, Neill, Newbry, Nief, Nomerotski, Nordby,
  O’Connor, Oliver, Olivier, Olsen, O’Mullane, Ortiz, Osier, Owen, Pain,
  Palecek, Parejko, Parsons, Pease, Peterson, Peterson, Petravick, Petrick,
  Petry, Pierfederici, Pietrowicz, Pike, Pinto, Plante, Plate, Plutchak, Price,
  Prouza, Radeka, Rajagopal, Rasmussen, Regnault, Reil, Reiss, Reuter, Ridgway,
  Riot, Ritz, Robinson, Roby, Roodman, Rosing, Roucelle, Rumore, Russo, Saha,
  Sassolas, Schalk, Schellart, Schindler, Schmidt, Schneider, Schneider,
  Schoening, Schumacher, Schwamb, Sebag, Selvy, Sembroski, Seppala, Serio,
  Serrano, Shaw, Shipsey, Sick, Silvestri, Slater, Smith, Smith, Sobhani,
  Soldahl, Storrie-Lombardi, Stover, Strauss, Street, Stubbs, Sullivan,
  Sweeney, Swinbank, Szalay, Takacs, Tether, Thaler, Thayer, Thomas, Thornton,
  Thukral, Tice, Trilling, Turri, Berg, Berk, Vetter, Virieux, Vucina, Wahl,
  Walkowicz, Walsh, Walter, Wang, Wang, Warner, Wiecha, Willman, Winters,
  Wittman, Wolff, Wood-Vasey, Wu, Xin, Yoachim, \& Zhan}]{Ivezic_2019}
Ivezi{\'c}, Z., Kahn, S.~M., Tyson, J.~A., {et~al.} 2019, The Astrophysical
  Journal, 873, 111, \dodoi{10.3847/1538-4357/ab042c}

\bibitem[{Jenniskens(2006)}]{jenniskens2006}
Jenniskens, P. 2006, Meteor Showers and their Parent Comets (Cambridge, U.K.:
  Cambridge University Press), 790

\bibitem[{Jenniskens(2008)}]{JENNISKENS200813}
---. 2008, Icarus, 194, 13,
  \dodoi{https://doi.org/10.1016/j.icarus.2007.09.016}

\bibitem[{Jenniskens(2023)}]{Jenniskens2023}
---. 2023, Atlas of Earth's Meteor Showers (Amsterdam, Netherlands: Elsevier
  Science), 838

\bibitem[{Jenniskens {et~al.}(2011)Jenniskens, Gural, Dynneson, Grigsby,
  Newman, Borden, Koop, \& Holman}]{cams}
Jenniskens, P., Gural, P., Dynneson, L., {et~al.} 2011, Icarus, 216, 40,
  \dodoi{https://doi.org/10.1016/j.icarus.2011.08.012}

\bibitem[{Jenniskens {et~al.}(2020)Jenniskens, Lyytinen, Johannink, Odeh,
  Moskovitz, \& Abbott}]{JENNISKENS2020104829}
Jenniskens, P., Lyytinen, E., Johannink, C., {et~al.} 2020, Planetary and Space
  Science, 181, 104829, \dodoi{https://doi.org/10.1016/j.pss.2019.104829}

\bibitem[{Jenniskens {et~al.}(2021)Jenniskens, Lauretta, Towner, Heathcote,
  Jehin, Hanke, Cooper, Baggaley, Howell, Johannink, Breukers, Odeh, Moskovitz,
  Juneau, Beck, {De Cicco}, Samuels, Rau, Albers, \& Gural}]{ms_lpc}
Jenniskens, P., Lauretta, D.~S., Towner, M.~C., {et~al.} 2021, Icarus, 365,
  114469, \dodoi{https://doi.org/10.1016/j.icarus.2021.114469}

\bibitem[{Jewitt {et~al.}(2007)Jewitt, Chizmadia, Grimm, \&
  Prialnik}]{Jewitt_2007}
Jewitt, D., Chizmadia, L., Grimm, R., \& Prialnik, D. 2007, Protostars and
  Planets V, 863

\bibitem[{Jewitt {et~al.}(2021)Jewitt, Kim, Mutchler, Agarwal, Li, \&
  Weaver}]{Jewitt_2021}
Jewitt, D., Kim, Y., Mutchler, M., {et~al.} 2021, The Astronomical Journal,
  161, 188, \dodoi{10.3847/1538-3881/abe4cf}

\bibitem[{Jopek(1993)}]{JOPEK1993603}
Jopek, T.~J. 1993, Icarus, 106, 603,
  \dodoi{https://doi.org/10.1006/icar.1993.1195}

\bibitem[{Jopek {et~al.}(2008)Jopek, Rudawska, \& Bartczak}]{Jopek2008}
Jopek, T.~J., Rudawska, R., \& Bartczak, P. 2008, Earth, Moon, and Planets,
  102, 73, \dodoi{10.1007/s11038-007-9197-8}

\bibitem[{Knight {et~al.}(2023)Knight, Kokotanekova, \&
  Samarasinha}]{knight2023physical}
Knight, M.~M., Kokotanekova, R., \& Samarasinha, N.~H. 2023, Physical and
  Surface Properties of Comet Nuclei from Remote Observations.
\newblock \doarXiv{2304.09309}

\bibitem[{Lamy {et~al.}(2004)Lamy, Toth, Fernandez, \&
  Weaver}]{comet_nuclei_properties}
Lamy, P.~L., Toth, I., Fernandez, Y.~R., \& Weaver, H.~A. 2004, in Comets II,
  ed. M.~C. Festou, H.~U. Keller, \& H.~A. Weaver (Tucson, AZ: University of
  Arizona Press), 317--335

\bibitem[{{LSST Science Collaboration} {et~al.}(2009){LSST Science
  Collaboration}, Abell, Allison, Anderson, Andrew, Angel, Armus, Arnett,
  Asztalos, Axelrod, Bailey, Ballantyne, Bankert, Barkhouse, Barr, Barrientos,
  Barth, Bartlett, Becker, Becla, Beers, Bernstein, Biswas, Blanton, Bloom,
  Bochanski, Boeshaar, Borne, Bradac, Brandt, Bridge, Brown, Brunner, Bullock,
  Burgasser, Burge, Burke, Cargile, Chandrasekharan, Chartas, Chesley, Chu,
  Cinabro, Claire, Claver, Clowe, Connolly, Cook, Cooke, Cooray, Covey,
  Culliton, de~Jong, de~Vries, Debattista, Delgado, Dell'Antonio, Dhital,
  Di~Stefano, Dickinson, Dilday, Djorgovski, Dobler, Donalek, Dubois-Felsmann,
  Durech, Eliasdottir, Eracleous, Eyer, Falco, Fan, Fassnacht, Ferguson,
  Fernandez, Fields, Finkbeiner, Figueroa, Fox, Francke, Frank, Frieman,
  Fromenteau, Furqan, Galaz, Gal-Yam, Garnavich, Gawiser, Geary, Gee, Gibson,
  Gilmore, Grace, Green, Gressler, Grillmair, Habib, Haggerty, Hamuy, Harris,
  Hawley, Heavens, Hebb, Henry, Hileman, Hilton, Hoadley, Holberg, Holman,
  Howell, Infante, Ivezic, Jacoby, Jain, Jedicke, Jee, Jernigan, Jha, Johnston,
  Jones, Juric, Kaasalainen, Kafka, Kahn, Kaib, Kalirai, Kantor, Kasliwal,
  Keeton, Kessler, Knezevic, Kowalski, Krabbendam, Krughoff, Kulkarni, Kuhlman,
  Lacy, Lepine, Liang, Lien, Lira, Long, Lorenz, Lotz, Lupton, Lutz, Macri,
  Mahabal, Mandelbaum, Marshall, May, McGehee, Meadows, Meert, Milani, Miller,
  Miller, Mills, Minniti, Monet, Mukadam, Nakar, Neill, Newman, Nikolaev,
  Nordby, O'Connor, Oguri, Oliver, Olivier, Olsen, Olsen, Olszewski, Oluseyi,
  Padilla, Parker, Pepper, Peterson, Petry, Pinto, Pizagno, Popescu, Prsa,
  Radcka, Raddick, Rasmussen, Rau, Rho, Rhoads, Richards, Ridgway, Robertson,
  Roskar, Saha, Sarajedini, Scannapieco, Schalk, Schindler, Schmidt, Schmidt,
  Schneider, Schumacher, Scranton, Sebag, Seppala, Shemmer, Simon, Sivertz,
  Smith, Smith, Smith, Spitz, Stanford, Stassun, Strader, Strauss, Stubbs,
  Sweeney, Szalay, Szkody, Takada, Thorman, Trilling, Trimble, Tyson, Van~Berg,
  Berk, VanderPlas, Verde, Vrsnak, Walkowicz, Wandelt, Wang, Wang, Warner,
  Wechsler, West, Wiecha, Williams, Willman, Wittman, Wolff, Wood-Vasey,
  Wozniak, Young, Zentner, \& Zhan}]{lsst_science_book}
{LSST Science Collaboration}, Abell, P.~A., Allison, J., {et~al.} 2009, LSST
  Science Book, Version 2.0,  arXiv, \dodoi{10.48550/ARXIV.0912.0201}

\bibitem[{{LSST Solar System Science Collaboration} {et~al.}(2020){LSST Solar
  System Science Collaboration}, Jones, Bannister, Bolin, Chandler, Chesley,
  Eggl, Greenstreet, Holt, Hsieh, Ivezić, Jurić, Kelley, Knight, Malhotra,
  Oldroyd, Sarid, Schwamb, Snodgrass, Solontoi, \&
  Trilling}]{science_impact_LSST_SSscience}
{LSST Solar System Science Collaboration}, Jones, R.~L., Bannister, M.~T.,
  {et~al.} 2020, The Scientific Impact of the Vera C. Rubin Observatory's
  Legacy Survey of Space and Time (LSST) for Solar System Science.
\newblock \doarXiv{2009.07653}

\bibitem[{Mainzer {et~al.}(2011)Mainzer, Grav, Bauer, Masiero, McMillan, Cutri,
  Walker, Wright, Eisenhardt, Tholen, Spahr, Jedicke, Denneau, DeBaun, Elsbury,
  Gautier, Comillion, Hand, Mo, J.Watkins, Wilkins, Bryngelson, Molina, Desai,
  Camus, Hidalgo, Konstantopoulos, Larsen, Maleszewski, Malkan, Mauduit,
  Mullan, Olszewski, Pforr, Saro, Scotti, \& Wasserman}]{neo_results}
Mainzer, A., Grav, T., Bauer, J., {et~al.} 2011, The Astrophysical Journal,
  743, 156–173

\bibitem[{Marsden \& Steel(1994)}]{Marsden_ihca}
Marsden, B., \& Steel, D. 1994, Warning Times and Impact Probablilities for
  Long-Period Comets (Tucson, AZ: University of Arizona Press), 221--239,
  \dodoi{https://doi.org/10.2307/j.ctv23khmpv}

\bibitem[{Meech \& Svoren(2004)}]{comet_nuclei}
Meech, K.~J., \& Svoren, J. 2004, in Comets II, ed. M.~C. Festou, H.~U. Keller,
  \& H.~A. Weaver (Tucson, AZ: University of Arizona Press), 317--335

\bibitem[{{Minor Planet Center}(2023)}]{mpc_nishimura}
{Minor Planet Center}. 2023, Minor Planet Circulars 164779.
\newblock
  \url{https://minorplanetcenter.net/iau/ECS/MPCArchive/2023/MPC_20230912.pdf}

\bibitem[{Morrison(2006)}]{Morrison2006}
Morrison, D. 2006, The Contemporary Hazard of Comet Impacts (Berlin,
  Heidelberg: Springer Berlin Heidelberg), 285--302,
  \dodoi{10.1007/3-540-33088-7_9}

\bibitem[{Nishimura {et~al.}(2023)Nishimura, Yoshimoto, Masek, Deen, Pearce,
  Sato, Takahashi, Mattiazzo, Ivanov, Banfalvy, Jaeger, Masi, Haver, Sicoli,
  Pettarin, Gonzalez, Paul, Balam, Green, \& Spratt}]{mpec_nishimura}
Nishimura, H., Yoshimoto, K., Masek, M., {et~al.} 2023, COMET C/2023 P1
  (Nishimura), Minor Planet Electronic Circular, No. 2023-P87,
  \dodoi{10.48377/MPEC/2023-P87}

\bibitem[{Nuth {et~al.}(2018)Nuth, Barbee, \& Leung}]{Nuth_2018}
Nuth, J.~A., Barbee, B., \& Leung, R. 2018, The Journal of Space Safety
  Engineering, 5(3-4), 197–202,
  \dodoi{https://doi.org/10.1016/j.jsse.2018.07.002}

\bibitem[{Pilorz {et~al.}(2023)Pilorz, Jenniskens, \& Vaubillon}]{Pilorz_2023}
Pilorz, S., Jenniskens, P., \& Vaubillon, J. 2023, in Asteroids, Comets,
  Meteors Conference 2023, LPI Contrib. No. 2851, 2484

\bibitem[{Quintana \& Schultz(2019)}]{QUINTANA2019176}
Quintana, S.~N., \& Schultz, P.~H. 2019, Icarus, 326, 176,
  \dodoi{https://doi.org/10.1016/j.icarus.2019.02.029}

\bibitem[{Rożek {et~al.}(2011)Rożek, Breiter, \& Jopek}]{Rozek_2011}
Rożek, A., Breiter, S., \& Jopek, T.~J. 2011, Monthly Notices of the Royal
  Astronomical Society, 412, 987, \dodoi{10.1111/j.1365-2966.2010.17967.x}

\bibitem[{Rudawska {et~al.}(2015)Rudawska, Matlovic, Tóth, \&
  Kornoš}]{Rudawska_2015}
Rudawska, R., Matlovic, P., Tóth, J., \& Kornoš, L. 2015, Proceedings of the
  International Meteor Conference, Mistelbach, Austria, 185

\bibitem[{Southworth \& Hawkins(1963)}]{s&h_d_criterion}
Southworth, R.~B., \& Hawkins, G.~S. 1963, Smithsonian Contributions to
  Astrophysics, 7, 261

\bibitem[{Tomko \& Neslusan(2019)}]{2p_encke}
Tomko, D., \& Neslusan, L. 2019, Astronomy \& Astrophysics, 623, A13,
  \dodoi{10.1051/0004-6361/201833868}

\bibitem[{Toon {et~al.}(1994)Toon, Zahnle, Turco, \& Covey}]{Toon_ihca}
Toon, O.~B., Zahnle, K., Turco, R.~P., \& Covey, C. 1994, Environmental
  Perturbations Caused by Asteroid Impacts (Tucson, AZ: University of Arizona
  Press), 791--826, \dodoi{https://doi.org/10.2307/j.ctv23khmpv}

\bibitem[{Valsecchi {et~al.}(1999)Valsecchi, Jopek, \&
  Froeschlé}]{Valsecchi_MSI_1_1999}
Valsecchi, G.~B., Jopek, T.~J., \& Froeschlé, C. 1999, Monthly Notices of the
  Royal Astronomical Society, 304, 743,
  \dodoi{10.1046/j.1365-8711.1999.02264.x}

\bibitem[{{Vaubaillon} {et~al.}(2005){Vaubaillon}, {Colas}, \&
  {Jorda}}]{Vaubaillon_2005}
{Vaubaillon}, J., {Colas}, F., \& {Jorda}, L. 2005, \aap, 439, 751,
  \dodoi{10.1051/0004-6361:20041544}

\bibitem[{Weissman(2006)}]{Weissman_2006}
Weissman, P.~R. 2006, Proceedings of the International Astronomical Union, 2,
  441–450, \dodoi{10.1017/S1743921307003559}

\bibitem[{Weryk(2023)}]{CBET_weryk}
Weryk, R. 2023, CBET \#5291, Central Bureau for Astronomical Telegrams, Hoffman
  Lab 209 Harvard University, 20 Oxford St., Cambridge, MA 02138, U.S.A.
\newblock \url{http://www.cbat.eps.harvard.edu/iau/cbet/005200/CBET005291.txt}

\bibitem[{Whipple(1950)}]{Whipple_1950}
Whipple, F.~L. 1950, Astrophysical Journal, 111, 375

\bibitem[{{Whitmell}(1907)}]{planet_brightness}
{Whitmell}, C.~T. 1907, The Observatory, 30, 96

\bibitem[{Ye(2016)}]{ye2016}
Ye, Q. 2016, Electronic Thesis and Dissertation Repository, 27.
\newblock \url{https://ir.lib.uwo.ca/etd/3903}

\bibitem[{Ye \& Greaves(2023)}]{CBET_ye}
Ye, Q., \& Greaves, J. 2023, CBET \#5290, Central Bureau for Astronomical
  Telegrams, Hoffman Lab 209 Harvard University, 20 Oxford St., Cambridge, MA
  02138, U.S.A.
\newblock \url{http://www.cbat.eps.harvard.edu/iau/cbet/005200/CBET005290.txt}

\bibitem[{Zhai {et~al.}(2020)Zhai, Ye, Shao, Trahan, Saini, Shen, Prince,
  Bellm, Graham, Helou, Kulkarni, Kupfer, Laher, Mahabal, Masci, Rusholme,
  Rosnet, \& Shupe}]{Zhai_2020}
Zhai, C., Ye, Q., Shao, M., {et~al.} 2020, Publications of the Astronomical
  Society of the Pacific, 132, 064502, \dodoi{10.1088/1538-3873/ab828b}

\bibitem[{Zubović {et~al.}(2015)Zubović, Vida, Gural, \& Šegon}]{gmn}
Zubović, D., Vida, D., Gural, P., \& Šegon, D. 2015, Proceedings of the
  International Meteor Conference, Mistelbach, Austria, 27

\bibitem[{Šegon {et~al.}(2017)Šegon, Vaubaillon, Gural, Vida, Andreić,
  Korlević, \& Skokić}]{Segon_2017}
Šegon, D., Vaubaillon, J., Gural, P.~S., {et~al.} 2017, A\&A, 598, A15,
  \dodoi{10.1051/0004-6361/201629100}

\end{thebibliography}
\bibliographystyle{aasjournal}



\end{document}